\documentclass[%
onecolumn,
nobalancelastpage, 
superscriptaddress,
amsmath,amssymb,
pra,
floatfix,
]{revtex4-2}

\usepackage{physics}
\usepackage{graphicx}
\usepackage{bm}
\usepackage{bbm} 
\usepackage{xcolor}
\usepackage{import}
\usepackage{siunitx}
\usepackage{hyperref}

\AtBeginDocument{\RenewCommandCopy\qty\SI}

\begin{document}

\title{Conditional Enhancement of Radical Pair Dynamics via Chiral State Preparation}

\author{Tristen Gwynn}
\email{tristen.gwynn@nithecs.ac.za}
\affiliation{Department of Physics, Stellenbosch University, Stellenbosch, South Africa}

\author{Betony Adams}
\affiliation{Department of Physics, Stellenbosch University, Stellenbosch, South Africa}
\affiliation{The Guy Foundation, Beaminster, Dorset, UK}

\author{Francesco Petruccione}
\affiliation{Department of Physics, Stellenbosch University, Stellenbosch, South Africa}
\affiliation{School for Data Science and Computational Thinking, Stellenbosch University,Stellenbosch, South Africa}
\affiliation{National Institute for Theoretical and Computational Sciences (NITheCS),Stellenbosch, South Africa}

\date{\today}
\begin{abstract}
Chiral-induced spin selectivity (CISS) has been shown to enhance magnetic sensitivity in radical pair mechanism (RPM) models under specific Hamiltonian conditions, yet whether these enhancements persist across a broader parameter space remains untested. We incorporate the CISS effect as a spin-dependent initial state and recombination operator and systematically evaluate the spin dynamics of a model radical pair across a comprehensive parameter sweep of the RPM Hamiltonian. We characterise the orientational response through symmetric and antisymmetric decomposition of the yield distribution under field reversal, providing a direct quantitative signature of CISS-induced symmetry breaking. Our analysis demonstrates that CISS does not function as a generic amplifier of magnetic sensitivity. Claimed enhancements are conditional on the relative alignment of the internal hyperfine and dipolar interaction axes, arising specifically under conditions of non-collinear internal interactions. Extension to a two-nucleus model confirms that these enhancements are sensitive to nuclear spin. CISS-induced effects observed in the single-nucleus model are substantially suppressed when a second collinear nucleus is introduced, with the exception of the hyperfine axis rotation sweep where non-collinear tensor misalignment drives a robust antisymmetric response. These findings indicate that the conditions for CISS-enhanced magnetoreception are more stringent than previously demonstrated, requiring highly ordered and rigid molecular geometries to sustain the effect. 
\end{abstract}
\maketitle

\section{Introduction}
\label{sec:intro}

The interaction between photo-excited electron radical pairs and environmental magnetic fields is widely proposed as the light-dependent mechanism underlying magnetoreception in avian photo-sensitive proteins~\cite{ritz2000model, hore2016radical}. Within this framework, Zeeman coupling with the Earth's magnetic field, together with exchange, dipolar, and hyperfine couplings, governs the spin dynamics that determine spin-selective reaction pathways, forming the basis of the radical pair mechanism (RPM)~\cite{lewis2018spin}.  Theoretical studies on the orientational properties of the RPM have demonstrated strong magnetic dependencies and `quantum needle' precision, consistent with experimental observations~\cite{bezchastnov2026arrangement,hiscock2016quantum,ren2021angular}.

Since these reactions occur within chiral protein environments, an additional spin selectivity may arise from molecular chirality~\cite{naaman2019chiral, bloom2024chiral}. The chiral-induced spin selectivity (CISS) effect describes the preferential transmission of electron spins through chiral structures~\cite{aiello2022chirality,naaman2012chiral}, with spin selectivity intrinsically linked to spatial handedness rather than chemical composition~\cite{naaman2020chiral, evers2022theory}. While theoretical descriptions commonly invoke spin–orbit coupling in chiral potentials, the microscopic origin of CISS remains an open question~\cite{bloom2024chiral,dalum2019theory, evers2022theory, fransson2022chiral, zollner2020insight}.

The introduction of a spin-selective process into the spin-sensitive RPM has motivated considerable interest~\cite{luo2021chiral, tiwari2022role}. Incorporating CISS into RPM models has been shown to enhance quantum coherence and magnetic sensitivity under certain conditions~\cite{fay2021chirality, tiwari2022role, tiwari2023quantum}, a finding further supported by recent direct spectroscopic evidence of CISS in radical-pair systems~\cite{latawiec2025detecting}. However, these enhancements are typically demonstrated for specific, isolated Hamiltonian parameter choices. Given that the magnetic response is highly sensitive to the interplay of internal interactions, it remains unclear whether CISS constitutes a robust, universal amplifier of magnetic sensitivity or whether its efficacy is bounded by strict structural constraints.

In this work, we address this question by systematically exploring the RPM Hamiltonian across physically relevant parameter ranges to characterise the conditions under which CISS-induced enhancement arises. To isolate the impact of spin selectivity without committing to a specific microscopic origin, we adopt the kinetic model of Luo and Hore~\cite{luo2021chiral}, treating CISS as a kinetic filter on the initial spin composition rather than a modification of the Hamiltonian dynamics. Section~\ref{sec:method} details the radical pair model, the Lindblad master equation framework, and the shelving state approach used to extract asymptotic reaction yields. Section~\ref{sec:results} presents a systematic evaluation of each internal interaction across the relevant parameter space, characterising how each modulates the orientational anisotropy and spatial symmetry of the reaction yields under chiral and non-chiral initial conditions. Section~\ref{sec:discussion} contextualises these findings within the broader framework of biological magnetoreception and examines the structural and dynamical conditions under which CISS may play a functional role.
\section{Model and Methods}
\label{sec:method}

In the absence of CISS, radical pairs are typically formed in the singlet state due to spin conservation and Pauli exclusion~\cite{hore2016radical},
\begin{equation}
        \ket{\psi_S} = \frac{1}{\sqrt{2}}( \ket{\alpha_D \beta_A} - \ket{\beta_D \alpha_A} ),
\end{equation}
where $\alpha$ and $\beta$ denote spin-up and spin-down states, and $D$ and $A$ label the donor and acceptor sites.

Within the Luo–Hore framework, CISS is incorporated as a spin-selective constraint on the electron transfer process, such that forward transfer and recombination favour opposite spin orientations~\cite{luo2021chiral}. This modifies the initially prepared spin state to,
\begin{equation}
\label{eq:initial_state}
        \ket{\psi_I} = \cos{(\chi/2)}\ket{\psi_S} + \sin{(\chi/2)}\ket{\psi_{T_0}},
\end{equation}
where the mixing angle $\chi \in [0, \pi/2]$ parametrises the strength of spin selectivity, and the triplet component $\ket{\psi_{T_0}}$ is defined as,
\begin{equation}
        \ket{\psi_{T_0}} = \frac{1}{\sqrt{2}}( \ket{\alpha_D \beta_A} + \ket{\beta_D \alpha_A} ).
\end{equation}
The corresponding chiral recombination state is,
\begin{equation}
\label{eq:recombination_state}
        \ket{\psi_R} = \cos{(\chi/2)}\ket{\psi_S} - \sin{(\chi/2)}\ket{\psi_{T_0}},
\end{equation}
ensuring the same parameter $\chi$ governs both state preparation and recombination. In this study, we contrast the limits of zero chirality ($\chi = 0$) and maximal CISS-induced triplet mixing ($\chi = \pi/2$). These two-electron states of $D$ and $A$ form the basis of the radical pair. However, in order to consider the spin evolution of the system we need to include the spin states of the nuclei. The initial density matrix is then constructed by averaging over all nuclear spin configurations,
\begin{equation}
    \rho_0 = \frac{1}{N} \sum_i \ket{\psi_{I,i}}\bra{\psi_{I,i}},
\end{equation}
where $N$ is the nuclear multiplicity corresponding to a thermal initial state.

The initial density matrix evolves under the total Hamiltonian~\cite{efimova2008role},
\begin{equation}
    \hat{H}_\text{Tot} = \hat{H}_\text{Z} + \hat{H}_\text{hf} + \hat{H}_\text{dip} + \hat{H}_\text{ex},
\end{equation}
where the Zeeman Hamiltonian is given by,
\begin{equation}
    \hat{H}_\text{Z}= -\gamma_e \mathbf{B} \cdot (\hat{\mathbf{S}}_D + \hat{\mathbf{S}}_A),
\end{equation}
where $\gamma_e$ is the gyromagnetic ratio of the electron, $\hat{\mathbf{S}}_i$ are the spin operators of the donor and acceptor electrons and the external magnetic field $\mathbf{B}$ can be expanded as,
\begin{equation}
    \mathbf{B}(B_0,\theta_\text{Z},\phi_\text{Z}) = B_0(\sin\theta_\text{Z} \cos\phi_\text{Z}, \sin\theta_\text{Z} \sin\phi_\text{Z},\cos\theta_\text{Z}),
\end{equation}
where $B_0$ is the field strength and $\theta_\text{Z}$ and $\phi_\text{Z}$ are the spherical coordinates of the magnetic field direction, illustrated by Fig.~\ref{fig:orientation_combined}. The hyperfine Hamiltonian is expressed as,
\begin{equation}
    \hat{H}_\text{hf}= \sum_{i\in D,A}\sum_k\hat{\mathbf{S}}_i \cdot \mathbf{A}_{ik} \cdot \hat{\mathbf{I}}_{ik},
\end{equation}
where $\hat{\mathbf{I}}_{ik}$ denotes the $k$-th nuclear spin operator associated with radical $i$. The hyperfine interaction tensor $\mathbf{A}_{ik}$ is decomposed as 
\begin{equation}
    \mathbf{A}=a_0\mathbbm{1}+\mathbf{T},
\end{equation}
where the isotropic Fermi contact term $a_0$ is given by~\cite{lewis2018spin}, 
\begin{equation}
  a_0 = \frac{2\mu_0 \gamma_e\gamma_{\text{N}}}{3}|{\psi(0)}|^2,
\end{equation}
where $\mu_0$ is the permeability of free space, $\gamma_{e,N}$ are the the electron and nuclear gyromagnetic ratios and $|{\psi(0)}|^2$ denotes the non-zero spin density at the nucleus. The anisotropic tensor $\mathbf{T}$ represents the electron-nuclear dipolar interaction expressed by,
\begin{equation}
    T_{kj}(\mathbf{r_i}) = \frac{\mu_0}{4 \pi}\frac{\gamma_e \gamma_{\text{N}}}{r_i^3} (\delta_{kj} - 3n_k n_j),
\end{equation}
where $r_i$ represents the distance between the nucleus and the centre of the electron spin density, and $n_k, n_j$ are the components of the unit vector $\mathbf{n}$ directed along $r_i$ \cite{carrington1967introduction}.

The principal axis of the hyperfine tensor $\mathbf{A}$ is oriented at polar angle $\theta_{\text{hf}}$ and azimuthal angle $\phi_{\text{hf}}$ relative to the reference molecular $z$-axis, which coincides with both the dipolar principal axis and the CISS polarisation direction. In this study, $\phi_{\text{hf}} = 0$ is fixed, confining the hyperfine rotation to the $xz$-plane, and $\theta_{\text{hf}}$ is swept to evaluate the dependence of CISS-induced enhancement on axial alignment.

The electron–electron dipolar interaction Hamiltonian $\hat{H}_\text{dip}$ is described by,
\begin{equation}
    \hat{H}_\text{dip}=\hat{\mathbf{S}}_D \cdot \mathbf{D} \cdot\hat{\mathbf{S}}_A,
\end{equation}
with the tensor $\mathbf{D}$ receiving the same treatment as the anisotropic hyperfine tensor,
\begin{equation}
    D_{kj}(r_{DA}) = \frac{\mu_0}{4\pi} \frac{\gamma_e^2}{r_{DA}^3} (\delta_{kj} - 3n_k n_j),
\end{equation}
where $r_{DA}$ represents the distance between the radicals.  The axial dipolar parameter $D_0$ is defined as~\cite{efimova2008role,lewis2018spin},
\begin{equation}
    D_0=-\frac{3}{2}\frac{\mu_0}{4\pi} \frac{\gamma_e^2}{r_{DA}^3}.
\end{equation}
Aligning the inter-radical vector with the $z$-axis of the principal axis system yields a traceless tensor with the diagonal components $D_{xx} = D_{yy} = -\frac{2}{3}D_0$ and $D_{zz} = \frac{4}{3}D_0$.
The electron exchange interaction Hamiltonian is given by,
\begin{equation}
        \hat{H}_\text{ex} = - J \left(2\hat{\mathbf{S}}_D \cdot \hat{\mathbf{S}}_A +\frac{1}{2}\mathbbm{1}\right),
 \end{equation}
 which describes the isotropic spin-spin coupling arising from the overlap of the electron wave functions. The exchange coupling constant $J$ decays exponentially with distance according to~\cite{efimova2008role}, 
 \begin{equation}
     J = J_0 e^{-\beta r},
 \end{equation}
 where $J_0$ is the exchange strength at contact ($r=0$) and $\beta$ is the decay constant dictating the range of the interaction.

To establish a comprehensive phenomenological baseline we focus on the variables responsible for the interaction strengths represented in Table~\ref{tab:parameters}. These interaction parameters are considered across both positive and negative regimes. Physically, the sign of the isotropic Fermi contact term $a_0$ dictates the preferred relative spin alignment, with negative values favouring parallel configurations~\cite{carrington1967introduction}. 

The anisotropic hyperfine principal component $T_{zz}$ and the electron-electron dipolar coupling $D_0$ govern the geometry of the internal magnetic environment. Notably, the transverse hyperfine components are fixed ($T_{xx} = T_{yy} = \SI{-0.05}{\milli\tesla}$) to lift the energy degeneracies and facilitate `quantum-needle' like avoided crossings~\cite{hiscock2016quantum}. Thus the $T_{zz}$ sweep is not strictly traceless and introduces a concurrent isotropic energy shift. This shift is a known artefact of the fixed transverse components and is noted as a limitation of the present parametrisation. Furthermore, while the dipolar interaction physically requires ($D_0 < 0$)~\cite{smith2025chirality, efimova2008role}, extending the sweep into the positive regime allows the full parameter landscape to be mapped and confirms that the observed anisotropy structures are not artefacts of the sign convention. Finally, the isotropic exchange coupling $J$ acts as a pure scalar energy shift, modulating the energetic ordering of the singlet and triplet manifolds without introducing structural anisotropy~\cite{lewis2018spin, efimova2008role}.

\begin{table}[t]
\caption{\label{tab:parameters} Summary of baseline Hamiltonian parameters and sweep ranges. All 1D sweeps consist of 31 discrete steps. The external magnetic field angles ($\theta_Z, \phi_Z$) are evaluated over a $36 \times 36$ grid at every point.}
\begin{ruledtabular}
\begin{tabular}{ccc}
\textbf{Symbol} & \textbf{Baseline Value} & \textbf{Sweep Range} \\
\colrule
$B_0$ & $0.05$ mT & $[0, 0.1]$ mT \\
$a_0$ & $0$ mT & $[-1, 1]$ mT \\
$T_{zz}$ & $1.5$ mT & $[-1, 1]$ mT \\
$D_0$ & $-0.4$ mT & $[-1, 1]$ mT \\
$J$ & $0$ mT & $[-1, 1]$ mT \\
$\theta_{\text{hf}}$ & $\pi/4$ rad & $[0, \pi]$ rad \\
$\phi_{\text{hf}}$ & $0$ rad & Fixed \\
$T_{xx}, T_{yy}$ & $-0.05$ mT & Fixed \\
\end{tabular}
\end{ruledtabular}
\end{table}

\subsection{Open Quantum System Dynamics}
The coherent evolution and irreversible reaction kinetics of the density matrix $\rho(t)$ are described by,
\begin{equation}
    \frac{d\rho}{dt} = -\frac{i}{\hbar} [\hat{H}, \rho] + \sum_k \left( \hat{L}_k \rho \hat{L}_k^\dagger - \frac{1}{2} \{\hat{L}_k^\dagger \hat{L}_k, \rho\} \right).
\end{equation}

To track product accumulation without population recycling, the Hilbert space is expanded to include five orthogonal shelving states: $\ket{S}$, $\ket{T_{-}}$, $\ket{T_{0}}$, $\ket{T_{+}}$, and $\ket{R}$~\cite{gauger2011sustained}. Spin-selective recombination is implemented via Haberkorn-style collapse operators~\cite{haberkorn1976density, ivanov2010consistent}. For each nuclear configuration $i$, the recombination operators are,
\begin{equation}
    \hat{L}_{n,i} = \sqrt{k_n} \ket{n} \bra{\psi_{n,i}},
\end{equation}
where $n \in \{S, T_-, T_0, T_+, R\}$. The primary observables are the asymptotic reaction yields, representing the fractional population accumulated in each shelving state channel,
\begin{equation}
    \Phi_n = \operatorname{Tr}\{\hat{P}_n\rho(t_{\max})\},
\end{equation}
where $\hat{P}_n = \ket{n}\bra{n}$.

To quantify the magnetic response we consider orientational sensitivity as the yield range, $\Delta\Phi=\Phi_\text{max}-\Phi_\text{min}$, of the respective channel over the angular grid described in Section~\ref{sec:limitations}. To isolate the enhancement provided by CISS, we define the quantitative chiral enhancement metric,
\begin{equation}
    K_\Phi = \Delta{\Phi_{\chi=\pi/2}} - \Delta{\Phi_{\chi=0}}.
\end{equation}
This allows for a direct comparison of sensitivity enhancement across different Hamiltonian parameter regimes.

To characterise how CISS modifies the spatial symmetry of the yield distribution, we further decompose the orientational response into its symmetric and antisymmetric components under field reversal. The mean-subtracted yield is defined as,
\begin{equation}
    \overline{\Phi}_n(\theta, \phi) = \Phi_n(\theta, \phi) 
    - \langle \Phi_n \rangle,
\end{equation}
where $\langle \Phi_n \rangle$ denotes the spherical average over the full orientational grid. The symmetric ($+$) and antisymmetric ($-$) components under the antipodal map $(\theta, \phi) \mapsto (\pi - \theta, \phi + \pi)$ are then,
\begin{equation}
    \overline{\Phi}^{\pm}_{n,m}(\theta, \phi) = \frac{1}{2} 
    \left[ \overline{\Phi}_n(\theta, \phi) \pm 
    \overline{\Phi}_m(\pi-\theta, \phi+\pi) \right].
\end{equation}
The spread of each component across the orientational sphere provides a direct measure of its contribution to the anisotropic response,
\begin{equation}
    \Delta\overline{\Phi}_{n,m}^{\pm} = 
    \max_{\theta,\phi} \left( \overline{\Phi}^{\pm}_{n,m} \right) 
    - \min_{\theta,\phi} \left( \overline{\Phi}^{\pm}_{n,m} \right).
\end{equation}
In a standard RPM without CISS, the singlet and central triplet channels satisfy $\Phi_n(\mathbf{B}) = \Phi_n(-\mathbf{B})$ individually due to their even parity under field reversal, while the outer triplet sublevels obey the cross-symmetry $\Phi_{T_+}(\mathbf{B}) = \Phi_{T_-}(-\mathbf{B})$~\cite{sakurai2017modern}. Consequently, the antisymmetric components $\Delta\overline{\Phi}^-_{n,n}$ for $n \in \{S, T_0, R\}$ and $\Delta\overline{\Phi}^-_{T_+,T_-}$ vanish identically in the baseline RPM. A non-zero antisymmetric spread in any channel therefore provides a direct quantitative signature of CISS-induced field-reversal symmetry breaking~\cite{luo2021chiral,rodgers2009chemical}.

\subsection{Model Limitations and Parametric Evaluation} \label{sec:limitations}
We define specific model constraints to explore how chiral effects manifest within this system. These establish the methodology as a controlled physical baseline designed to isolate Hamiltonian effects, rather than a direct physiological simulation of an environment like cryptochrome.

First, the primary parametric evaluation is restricted to a single nuclear spin coupled to the donor electron. While biological radical pairs possess dense nuclear environments that drive decoherence, this minimal model prevents parameter degeneracy from obscuring the sensitivity structures of interest and isolates the geometric Hamiltonian effects that are the primary focus of this study. To assess which of these effects survive beyond the single-nucleus topology, the model is extended in Section~\ref{sec:results} to include a second spin-$\frac{1}{2}$ nucleus coupled to the acceptor radical, with its hyperfine axis fixed at the baseline orientation $\theta_\text{hf} = \pi/4$. This two-nucleus evaluation is computed at reduced angular resolution, $11 \times 11$, relative to the primary sweeps due to the increased computational cost of the larger Hilbert space, and serves as a robustness assessment rather than a full parametric evaluation. The corresponding symmetry decomposition plots are provided in Appendix~\ref{app:two_nuc}.

Second, to prevent kinetic asymmetries from artificially enhancing or suppressing sensitivity structures, all recombination rates are set equal ($k_S = k_T = k_R = k = \SI{0.1}{\per\micro\second}$). Recent studies, such as those by Smith \textit{et al.}~\cite{smith2025chirality}, demonstrate that CISS-induced polarisation can significantly bolster magnetic sensitivity under highly asymmetric reaction kinetics via the quantum Zeno effect. By setting equal rates, we deliberately decouple geometric alignment effects from Zeno-mediated kinetic dynamics, ensuring that any observed chiral enhancement is driven purely by internal tensor geometry.

In the baseline RPM ($\chi=0$), the singlet ($S$) and CISS-recombination ($R$) channels are degenerate~\cite{luo2021chiral}. Given our assignment of equal rate constants ($k_S = k_R$), the reaction yield is partitioned equally between the signalling pathway and the non-signalling return to the ground state.

\begin{figure}[htbp]
    \centering
    \includegraphics[width=\linewidth]{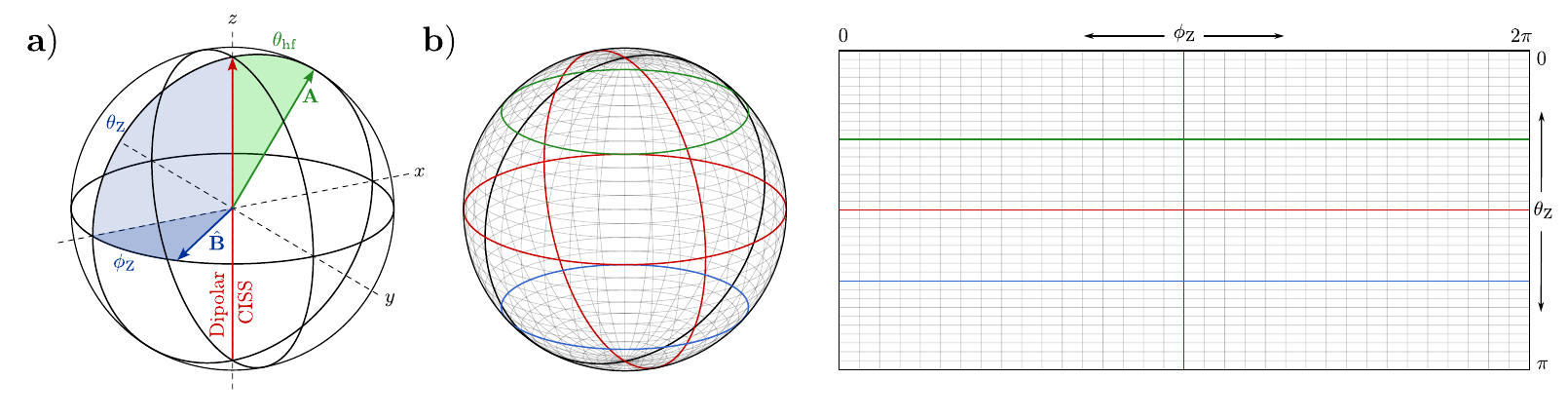}
    
    \caption{Coordinate system and discrete sampling of the external magnetic field orientation. \textbf{(a)} Spherical coordinate system defining the relative orientations of the radical pair Hamiltonian components. The $z$-axis is aligned with the inter-radical dipolar vector and CISS axis (red). The orientation of the external magnetic field, $\hat{\mathbf{B}}$, is defined by the Zeeman polar and azimuthal angles, $\theta_Z$ and $\phi_Z$ (blue). The principal axis of the hyperfine interaction (green) is oriented at an angle $\theta_{\text{hf}}$ relative to the molecular frame. \textbf{(b)} Flattened $36 \times 36$ rectangular grid of the orientational space, where $\theta_Z \in [0, \pi]$ and $\phi_Z \in [0, 2\pi]$ with step sizes of $\Delta\theta \approx 5.14^\circ$ and $\Delta\phi \approx 10.3^\circ$. Coloured lines on the sphere in (a) correspond to the respective longitudinal and latitudinal trajectories highlighted on the 2D grid in (b).}
    \label{fig:orientation_combined}
\end{figure}

The master equation is solved numerically using an implicit backward differentiation formula (BDF) integrator in QuTiP~\cite{johansson2012qutip}, evolving to $t_{\text{max}} = 5.0/k$. The code is made available at~\cite{gwynn2026ciss}. The dipolar tensor is strictly aligned with the laboratory reference frame ($\theta_D = 0, \phi_D = 0$) with principal values $D_{xx} = D_{yy} = -\frac{2}{3}D$ and $D_{zz} = \frac{4}{3}D$. For every parameter sweep point, the external magnetic field vector is sampled over a discrete orientational sphere and flattened to form a grid of $36 \times 36$ points spanning $\theta_Z \in [0,\pi]$ and $\phi_Z \in [0,2\pi]$ (Fig.~\ref{fig:orientation_combined}).

\section{Results}
\label{sec:results}

The following subsections evaluate how each Hamiltonian interaction modulates $K_\Phi$ and $\Delta\overline{\Phi}_{n}^{\pm}$. The external Zeeman field is examined first to establish the baseline orientational response and confirm the expected symmetry properties. The isotropic hyperfine, anisotropic hyperfine, and relative axis alignment are then evaluated to characterise the resonance and geometric conditions governing CISS-induced enhancement. The dipolar and exchange interactions are assessed subsequently, followed by a two-nucleus extension to test the robustness of the observed effects beyond the single-nucleus topology.

\subsection{Zeeman Interaction}
\label{sec:zeeman}

\begin{figure}[h]
    \centering
    \includegraphics[width=\linewidth]{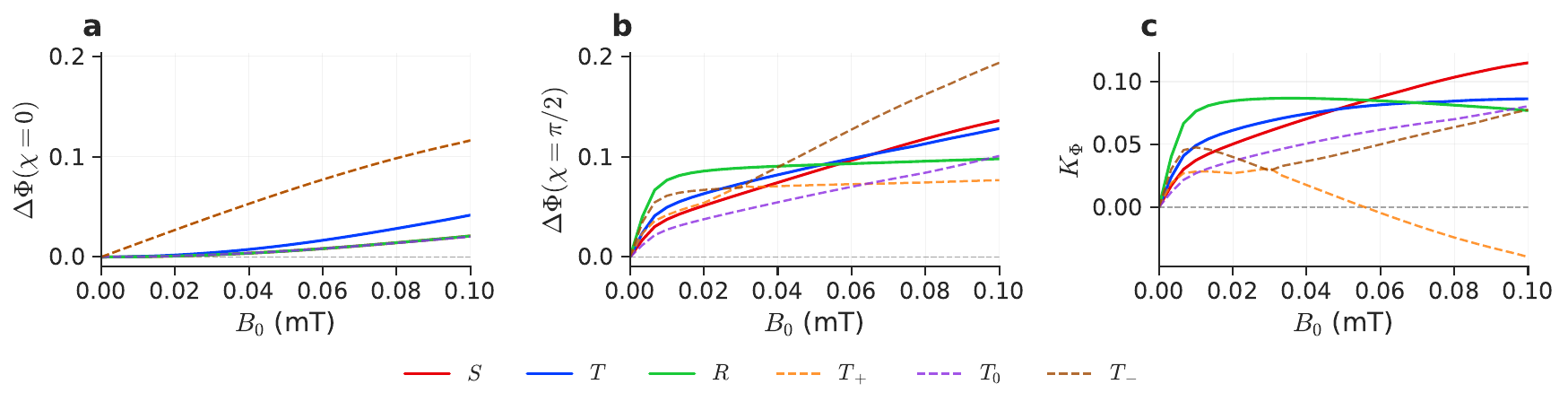}
    \caption{Orientational anisotropy and CISS-induced enhancement as a function of Zeeman coupling ($B_0$). Panels (a) and (b) show the yield range $\Delta\Phi$ across the full orientational grid for the baseline RPM ($\chi = 0$) and maximal CISS ($\chi = \pi/2$) respectively, for all spin channels. Panel (c) shows the chiral enhancement metric $K_\Phi$, isolating the net effect of CISS on orientational sensitivity. In panel (a), the $S$ and $R$ channels are degenerate and overlap; $T_+$ and $T_-$ are likewise degenerate in the baseline RPM and appear as a single line, a degeneracy lifted under CISS in panel (b). This degeneracy pattern appears in all subsequent anisotropy panels. A crossover near \SI{0.055}{\milli\tesla} marks the field at which the $S$ and $T$ channels surpass $R$ and $T_+$ in CISS-induced sensitivity, with $K_\Phi$ for $T_+$ becoming negative beyond this point.}
    \label{fig:anisotropy_B}
\end{figure}

The external Zeeman field ($B_0$) provides the primary orientational reference for the radical pair. At $B_0 = \SI{0}{\milli\tesla}$, yields are isotropic for both $\chi = 0$ and $\chi = \pi/2$, consistent with the absence of an external field. As the field increases, Zeeman splitting lifts the degeneracy of the triplet sublevels, separating the $T_\pm$ states from the $T_0$ and singlet manifolds and generating a growing orientational anisotropy in the baseline RPM (Fig.~\ref{fig:anisotropy_B}a).

Under CISS, the anisotropy increases sharply between 0 and \SI{0.01}{\milli\tesla} across all channels (Fig.~\ref{fig:anisotropy_B}b), a response absent in the baseline RPM where anisotropy grows gradually from zero. Beyond \SI{0.01}{\milli\tesla} the trends diverge: the $R$ and $T_+$ channels plateau and become insensitive to further increases in field strength, while the $S$, $T$, and $T_-$ channels continue to grow. By approximately \SI{0.055}{\milli\tesla}, the $S$ and total triplet $T$ channels exceed $R$ and $T_+$ in orientational sensitivity, and $K_\Phi$ for the $T_+$ channel becomes negative (Fig.~\ref{fig:anisotropy_B}c), indicating that CISS actively suppresses its anisotropy relative to the baseline RPM at higher fields. The differential field-dependence of the $R$ and $T_+$ channels beyond this regime is noted as an observation requiring further analysis.

\begin{figure}[h]
    \centering
    \includegraphics[width=\linewidth]{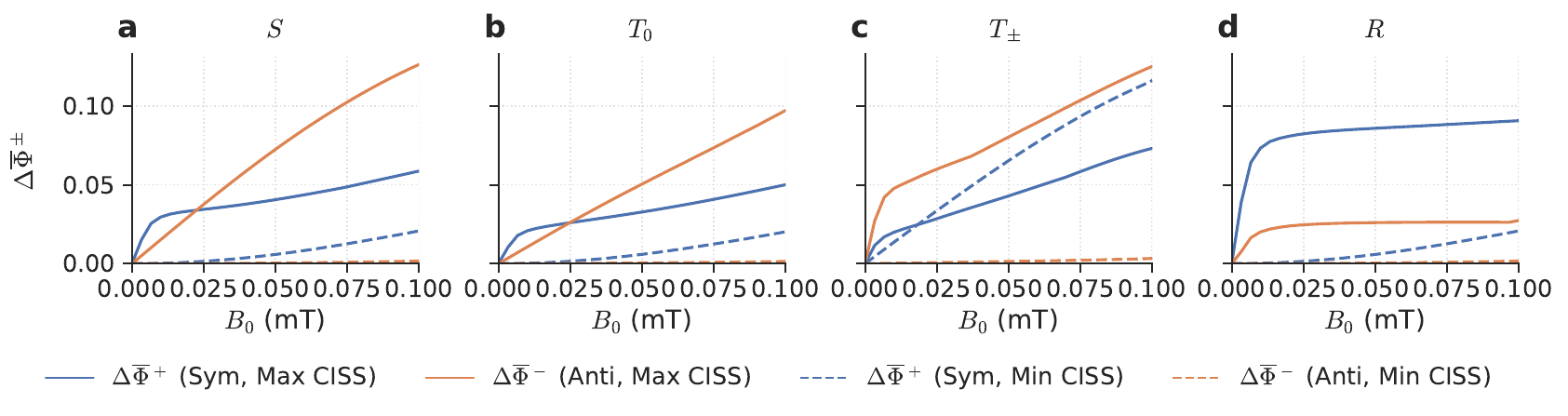}
    \caption{Symmetric ($\Delta\overline{\Phi}^+$) and antisymmetric ($\Delta\overline{\Phi}^-$) components of the orientational yield distribution under Zeeman coupling ($B_0$), shown for the $S$, $T_0$, $T_\pm$, and $R$ channels. Solid lines denote strong CISS ($\chi = \pi/2$); dashed lines denote the baseline RPM ($\chi = 0$). The antisymmetric component vanishes identically in the baseline RPM but acquires a significant, monotonically growing magnitude under strong CISS across all channels, constituting a direct signature of CISS-induced field-reversal symmetry breaking.}
    \label{fig:symmetry_B}
\end{figure}

The symmetric and antisymmetric decomposition (Fig.~\ref{fig:symmetry_B}) reveals the distinct contributions to this response. For the $S$, $T_0$, and $R$ channels, $\Delta\overline{\Phi}^+$ closely tracks the total anisotropy $\Delta\Phi$, confirming that the decomposition correctly partitions the orientational response, and grows monotonically under both chirality conditions with a steeper gradient under strong CISS. The field-reversal antisymmetric component $\Delta\overline{\Phi}^-$, which vanishes identically in the baseline RPM, acquires a significant monotonically increasing magnitude under strong CISS, most pronounced in $S$ and less so in $T_0$ and $R$. For the $T_\pm$ channels, the symmetric component is dampened under strong CISS while a comparable antisymmetric component is present. A small non-zero antisymmetric component also emerges in the baseline RPM with increasing $B_0$ in $T_\pm$, suggesting a weak intrinsic field-reversal asymmetry in the single-nucleus model that CISS substantially amplifies.

Taken together, the Zeeman results establish that CISS universally introduces a non-zero field-reversal antisymmetric yield component across all channels, growing with magnetic field strength. This provides the baseline against which the parameter-dependent modulation of both $K_\Phi$ and $\Delta\overline{\Phi}^-$ is assessed in the following subsections.

\subsection{Isotropic Hyperfine Interaction}
\label{sec:iso_hyperfine}

\begin{figure}[h]
    \centering
    \includegraphics[width=\linewidth]{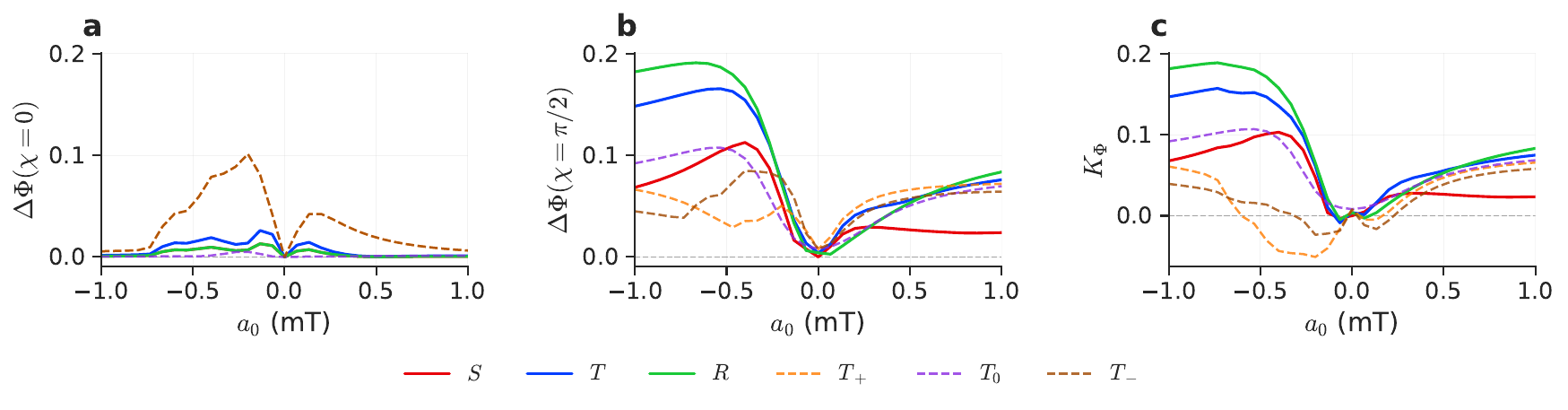}
    \caption{Orientational anisotropy and CISS-induced enhancement as a function of isotropic hyperfine coupling ($a_0$). Panels (a) and (b) show the yield range $\Delta\Phi$ for the baseline RPM ($\chi = 0$) and maximal CISS ($\chi = \pi/2$) respectively. Panel (c) shows $K_\Phi$, isolating the net CISS contribution to orientational sensitivity. For all channels except the outer triplet, $K_\Phi$ is positive across both signs of $a_0$. The $T_\pm$ channels show a reduction at low $|a_0|$, recovering to positive values at larger $|a_0|$, with this suppression more pronounced on the negative $a_0$ side.}
    \label{fig:anisotropy_Ai}
\end{figure}

In the absence of hyperfine coupling ($a_0 = \SI{0}{\milli\tesla}$), the system lacks a singlet--triplet mixing mechanism and the orientational anisotropy vanishes for both $\chi = 0$ and $\chi = \pi/2$ (Fig.~\ref{fig:anisotropy_Ai}a,b). As $|a_0|$ increases, the isotropic hyperfine field drives singlet--triplet interconversion, generating a growing orientational response. In the baseline RPM, the anisotropy peaks near $\pm\SI{0.2}{\milli\tesla}$ before declining, with the peak broader and more persistent for negative $a_0$ values, most visible in the outer triplet channels.

Under strong CISS, the anisotropy significantly increases relative to the baseline RPM, while preserving the near-zero response at $a_0 = \SI{0}{\milli\tesla}$. The asymmetry between positive and negative $a_0$ is strongly amplified (Fig.~\ref{fig:anisotropy_Ai}b), with negative values sustaining elevated anisotropy to approximately $\SI{-0.5}{\milli\tesla}$ before declining. The chiral enhancement metric $K_\Phi$ (Fig.~\ref{fig:anisotropy_Ai}c) reveals that this amplification is not uniform across channels. For the $S$, $T_0$, $T$, and $R$ channels, $K_\Phi$ increases for both signs of $a_0$, though more strongly for negative values. The outer triplet channels $T_\pm$ respond differently. For negative $a_0$, $K_\Phi$ initially decreases before recovering and becoming positive at larger magnitudes. For positive $a_0$, $K_\Phi$ is marginally negative at low values before recovering rapidly and increasing monotonically. This initial suppression before recovery is specific to the outer triplet channels. The preferential response to negative $a_0$ across all channels points to a role of the fixed negative dipolar coupling, characterised further in the two-dimensional map below.

\begin{figure*}[tb]
    \centering
    \includegraphics[width=\linewidth]{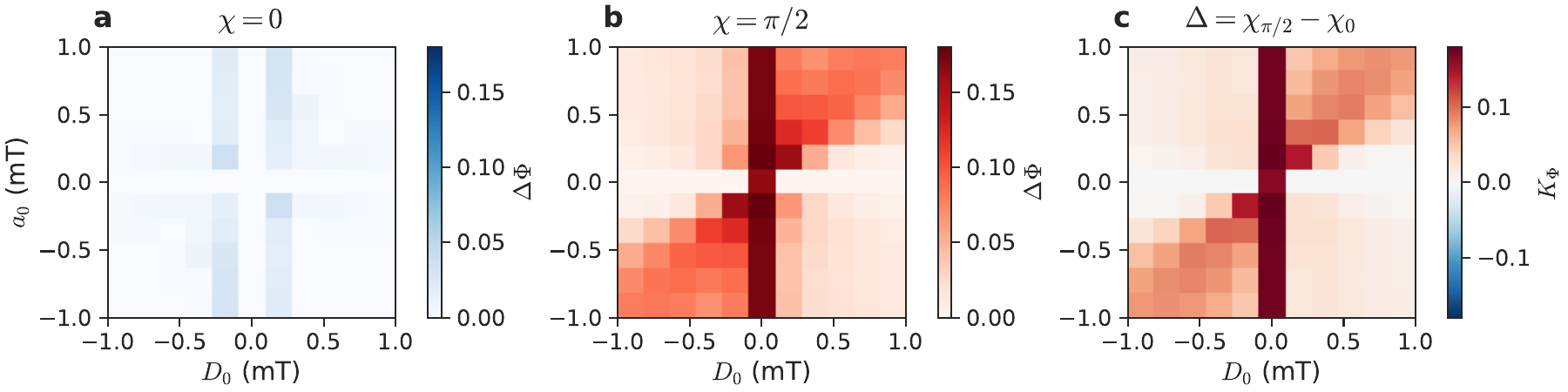}
    \caption{Two-dimensional map of orientational anisotropy $\Delta\Phi$ as a function of dipolar coupling ($D_0$) and isotropic hyperfine coupling ($a_0$), for (a) baseline RPM ($\chi = 0$), (b) maximal CISS ($\chi = \pi/2$), and (c) the net CISS-induced enhancement $K_\Phi$. In the baseline RPM, high anisotropy occurs where $a_0$ and $D_0$ carry opposite signs. Under CISS, this relationship inverts, with maximum enhancement concentrated along the same-sign diagonal. The highest CISS-induced anisotropy occurs near $D_0 = \SI{0}{\milli\tesla}$, a feature absent in the baseline RPM.}
    \label{fig:anisotropy_heatmap}
\end{figure*}

The combined influence of $a_0$ and $D_0$ on the CISS-induced enhancement is characterised in the two-dimensional map (Fig.~\ref{fig:anisotropy_heatmap}). In the baseline RPM, maximum anisotropy occurs at low interaction strengths where $a_0$ and $D_0$ carry opposite signs. As interaction strengths increase, this shifts toward the same-sign, while retaining the preference for low $D_0$. Under strong CISS, high anisotropy concentrates along the positive diagonal, with the same-sign regions showing significant enhancement relative to the baseline. The net enhancement $K_\Phi$ is therefore positive along the diagonal and suppressed in the opposite-sign regions. Additionally, the $D_0 = \SI{0}{\milli\tesla}$ limit shows the highest CISS-induced anisotropy across all $a_0$ values, a feature absent in the baseline RPM, indicating that the absence of dipolar coupling is a particularly favourable condition for CISS enhancement.

\begin{figure}[h]
    \centering
    \includegraphics[width=\linewidth]{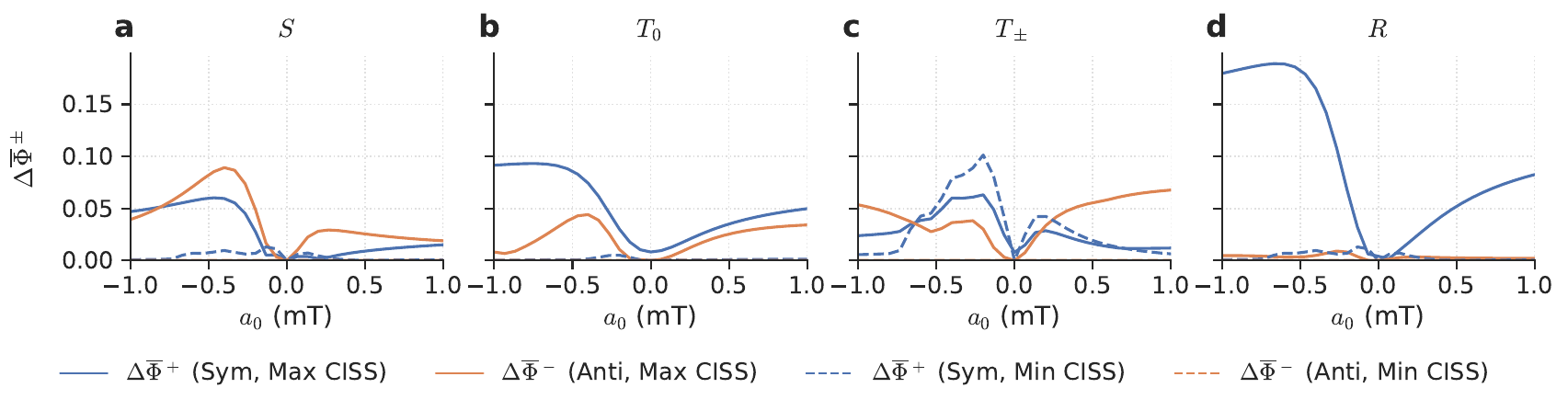}
    \caption{Symmetric ($\Delta\overline{\Phi}^+$) and antisymmetric ($\Delta\overline{\Phi}^-$) components of the orientational yield distribution as a function of isotropic hyperfine coupling ($a_0$), shown for the $S$, $T_0$, $T_\pm$, and $R$ channels. Solid lines denote strong CISS ($\chi = \pi/2$); dashed lines denote the baseline RPM ($\chi = 0$). The antisymmetric component is absent without CISS and grows under strong CISS with channel-dependent character: exceeding the symmetric in $S$, subdominant in $T_0$ and $T_\pm$, and negligible in $R$. A sharp decline in the antisymmetric component of $T_0$ beyond $\SI{-0.5}{\milli\tesla}$ is noted as a feature without a symmetric counterpart.}
    \label{fig:symmetry_Ai}
\end{figure}

The symmetric and antisymmetric decomposition (Fig.~\ref{fig:symmetry_Ai}) shows a pronounced bias toward negative $a_0$ across all channels, consistent with the anisotropy results. The field-reversal antisymmetric component $\Delta\overline{\Phi}^-$ is absent in the baseline RPM and emerges exclusively under strong CISS, with markedly different character between channels. In $S$, the antisymmetric component exceeds the symmetric for both signs of $a_0$. In $T_0$, the symmetric component dominates, with a sharp decline in the antisymmetric contribution beyond approximately $\SI{-0.5}{\milli\tesla}$ that has no symmetric counterpart. The outer triplet channels show reduced symmetric response for $|a_0| \lesssim \SI{0.5}{\milli\tesla}$, with an antisymmetric peak near $\pm\SI{0.25}{\milli\tesla}$ that is larger for positive $a_0$. The $R$ channel maintains a strongly symmetric response with negligible antisymmetric character throughout.

\subsection{Anisotropic Hyperfine Interaction}
\label{sec:aniso_hyperfine}

\begin{figure}[h]
    \centering
    \includegraphics[width=\linewidth]{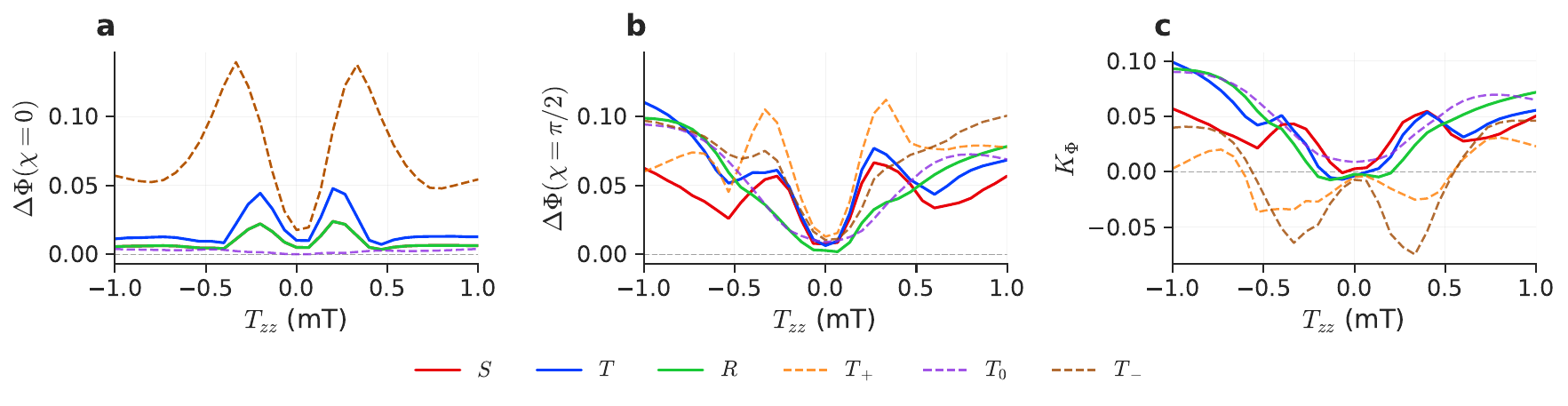}
    \caption{Orientational anisotropy and CISS-induced enhancement as a function of anisotropic hyperfine coupling ($T_{zz}$), evaluated at $\theta_\text{hf} = \pi/4$. Panels (a) and (b) show $\Delta\Phi$ for the baseline RPM ($\chi = 0$) and maximal CISS ($\chi = \pi/2$) respectively. Panel (c) shows $K_\Phi$. Distinct peaks near $\pm\SI{0.27}{\milli\tesla}$ are observed in the baseline RPM. Under CISS, the anisotropy continues to increase beyond $\pm\SI{0.5}{\milli\tesla}$ rather than plateauing, while $K_\Phi$ is positive for $S$, $T_0$, and $R$ at the peak values but negative for the outer triplet channels $T_\pm$.}
    \label{fig:anisotropy_Az}
\end{figure}

The anisotropic hyperfine interaction produces the most sharply peaked sensitivity structure observed across the parameter sweeps. The response is approximately mirror-symmetric about $T_{zz} = \SI{0}{\milli\tesla}$ at $\theta_\text{hf} = \pi/4$. The symmetry is not exact, since the fixed transverse components $T_{xx} = T_{yy} = \SI{-0.05}{\milli\tesla}$ and the fixed dipolar coupling prevent a $T_{zz}$ sign inversion from being a true symmetry operation. At $T_{zz} = \SI{0}{\milli\tesla}$, the orientational anisotropy reaches a clear local minimum but does not vanish, as the fixed transverse components retain a residual contribution that sustains weak spin mixing. As $|T_{zz}|$ increases, the anisotropy rises sharply to reach distinct peaks near $\pm\SI{0.27}{\milli\tesla}$ before falling away and plateauing for $|T_{zz}| \gtrsim \SI{0.5}{\milli\tesla}$ in the baseline RPM (Fig.~\ref{fig:anisotropy_Az}a). The $T_0$ channel shows a notably weak response throughout. The outer triplet channels $T_\pm$ exhibit the largest peaks, aligned with those of the other channels but considerably higher in magnitude, with their maxima appearing closer to $\pm\SI{0.4}{\milli\tesla}$.

Under strong CISS (Fig.~\ref{fig:anisotropy_Az}b), the double-peak structure is preserved and amplified in the same parameter region. In contrast to the baseline RPM where the anisotropy plateaus beyond $|T_{zz}| \gtrsim \SI{0.5}{\milli\tesla}$, the CISS case continues to increase monotonically with $|T_{zz}|$ across all channels.

The chiral enhancement metric $K_\Phi$ (Fig.~\ref{fig:anisotropy_Az}c) shows that this amplification is channel-dependent. For the $S$, $T_0$, and $R$ channels, $K_\Phi$ exhibits a double-lobe structure with positive peaks at the same $T_{zz}$ values as the anisotropy peaks. The outer triplet channels $T_\pm$ show the opposite response: $K_\Phi$ is negative at these peak locations, indicating CISS suppresses rather than amplifies the outer triplet anisotropy there. These results are obtained at $\theta_\text{hf} = \pi/4$; the dependence on hyperfine axis orientation is examined in the following subsection.

\begin{figure}[h]
    \centering
    \includegraphics[width=\linewidth]{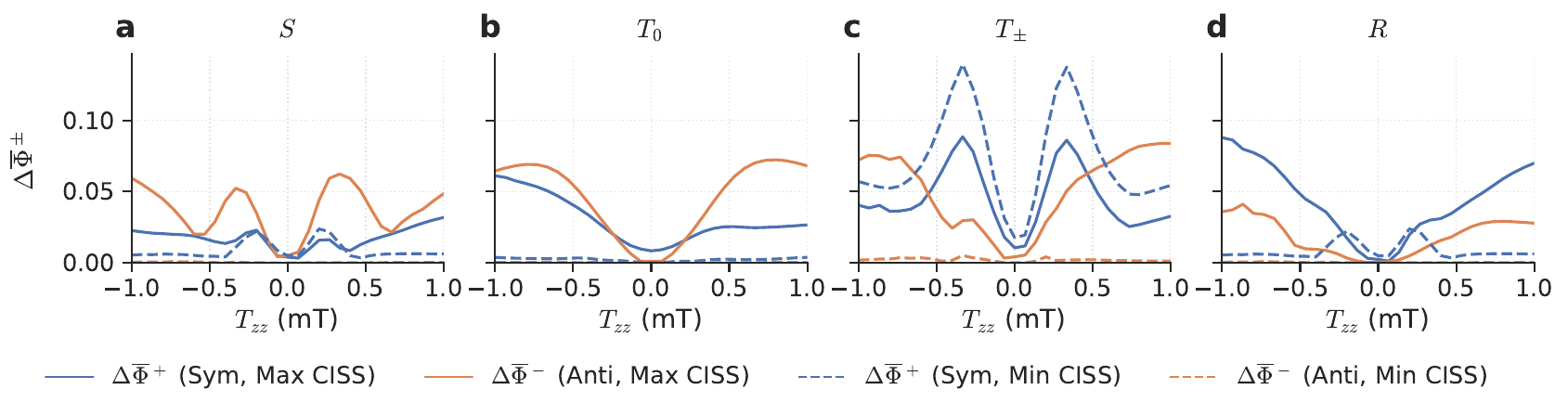}
    \caption{Symmetric ($\Delta\overline{\Phi}^{+}_S$) and antisymmetric ($\Delta\overline{\Phi}^{-}_S$) components of the orientational yield distribution as a function of anisotropic hyperfine coupling ($T_{zz}$), evaluated at $\theta_\text{hf} = \pi/4$, shown for $S$, $T_0$, $T_\pm$, and $R$. Under strong CISS, the symmetric peak of the singlet is suppressed while its antisymmetric component grows at both positive and negative $T_{zz}$ values. The outer triplet channels show CISS-induced suppression of their symmetric peaks. The $R$ channel displays a roughly symmetric response with a V-shaped antisymmetric profile.}
    \label{fig:symmetry_Az}
\end{figure}

The symmetric and antisymmetric decomposition (Fig.~\ref{fig:symmetry_Az}) reflects the double-lobe structure with channel-dependent character. For $S$, CISS suppresses the symmetric peak while producing a significant antisymmetric component at both positive and negative $T_{zz}$. The $T_0$ response shows a dominant antisymmetric component that is itself symmetric about $T_{zz} = \SI{0}{\milli\tesla}$, reaching zero at this point. The outer triplet channels show CISS-induced suppression of their symmetric peaks, with a broad V-shaped antisymmetric profile containing a small feature near $\SI{-0.27}{\milli\tesla}$. The $R$ channel displays a roughly symmetric response with an approximately monotonically growing antisymmetric V-shaped profile.

The anisotropic hyperfine results show peak sensitivity near $\pm\SI{0.27}{\milli\tesla}$, with CISS selectively amplifying these peaks for $S$, $T_0$, and $R$ while suppressing them for $T_\pm$. Beyond the peak region, CISS sustains a growing anisotropy where the baseline RPM saturates. The field-reversal antisymmetric component emerges under CISS across all channels, with character that varies between channel groups.

\subsection{Relative Axis Alignment}
\label{sec:axis_alignment}

The preceding subsections evaluated the anisotropic hyperfine interaction at a fixed orientation of $\theta_\text{hf} = \pi/4$. Here we examine how the orientation of the hyperfine principal axis relative to the shared dipolar and CISS $z$-axis governs the orientational response, through two complementary approaches: a sweep of $\theta_\text{hf}$ at fixed $T_{zz}$, and a sweep of $T_{zz}$ at three discrete angles.

\subsubsection*{Angular sweep at fixed \texorpdfstring{$T_{zz}$}{Tzz}}

\begin{figure}[h]
    \centering
    \includegraphics[width=\linewidth]{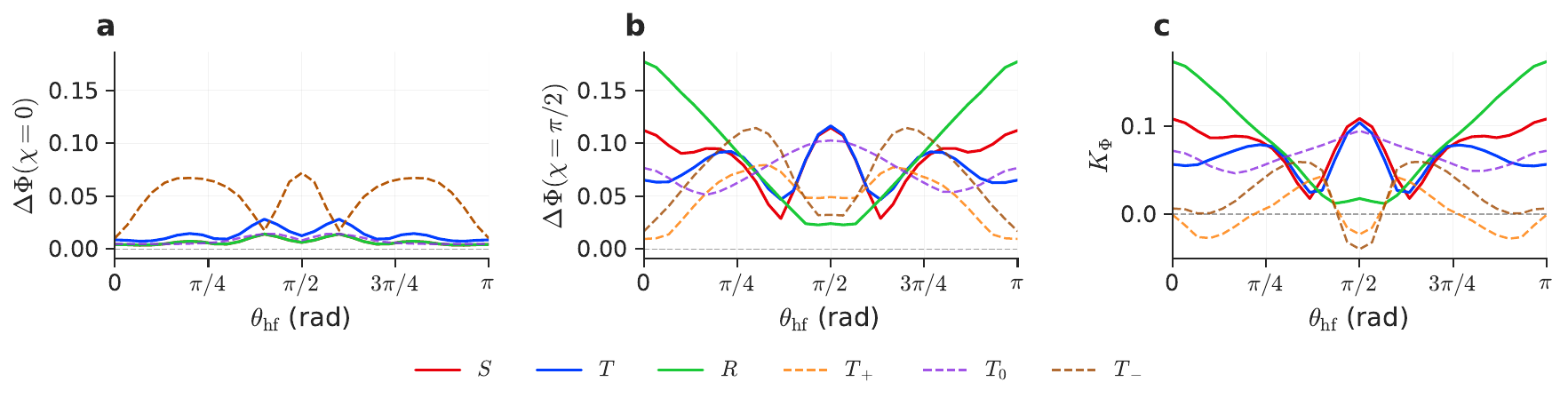}
    \caption{Orientational anisotropy and CISS-induced enhancement as a function of hyperfine orientation angle ($\theta_\text{hf}$) at fixed $T_{zz}$. Panels (a) and (b) show $\Delta\Phi$ for the baseline RPM ($\chi = 0$) and maximal CISS ($\chi = \pi/2$) respectively. Panel (c) shows $K_\Phi$. The outer triplet channels dominate the baseline RPM anisotropy with a response mirror-symmetric about $\pi/2$, while CISS redistributes sensitivity toward the $S$, $T_0$, and $R$ channels and converts the $\pi/2$ peak of the outer triplets into a local minimum.}
    \label{fig:anisotropy_theta_hf}
\end{figure}

Rotation of the hyperfine principal axis from parallel ($\theta_\text{hf} = 0$) to antiparallel ($\theta_\text{hf} = \pi$) reveals a clear angle-dependent orientational response (Fig.~\ref{fig:anisotropy_theta_hf}). In the baseline RPM, the outer triplet channels $T_\pm$ dominate the anisotropy and display a response mirror-symmetric about $\theta_\text{hf} = \pi/2$, with pronounced minima at $\theta_\text{hf} = 0$, $\pi$ and at approximately $2\pi/5$ and $3\pi/5$. These same angles correspond to maxima in the $S$, $T_0$, and $R$ channels, indicating a complementary redistribution of orientational sensitivity between channel groups as the hyperfine axis rotates.

Under strong CISS, this angular structure is substantially reshaped. The $R$ channel develops a clear monotonic relationship with alignment: anisotropy is greatest when the hyperfine axis is parallel to the CISS axis and lowest when antiparallel. The $S$ and total triplet $T$ channels show a strong response to the antiparallel configuration. The central peak at $\theta_\text{hf} = \pi/2$ present in the baseline RPM becomes a local minimum under CISS for the outer triplet channels, with their maxima shifting to intermediate angles. The enhancement metric $K_\Phi$ (Fig.~\ref{fig:anisotropy_theta_hf}c) confirms that CISS reduces the outer triplet anisotropy when the hyperfine axis is antiparallel, with additional minima near $2\pi/5$ and $3\pi/5$, while enhancing the $S$, $T_0$, and $R$ channels most strongly under antiparallel alignment.

\begin{figure}[h]
    \centering
    \includegraphics[width=\linewidth]{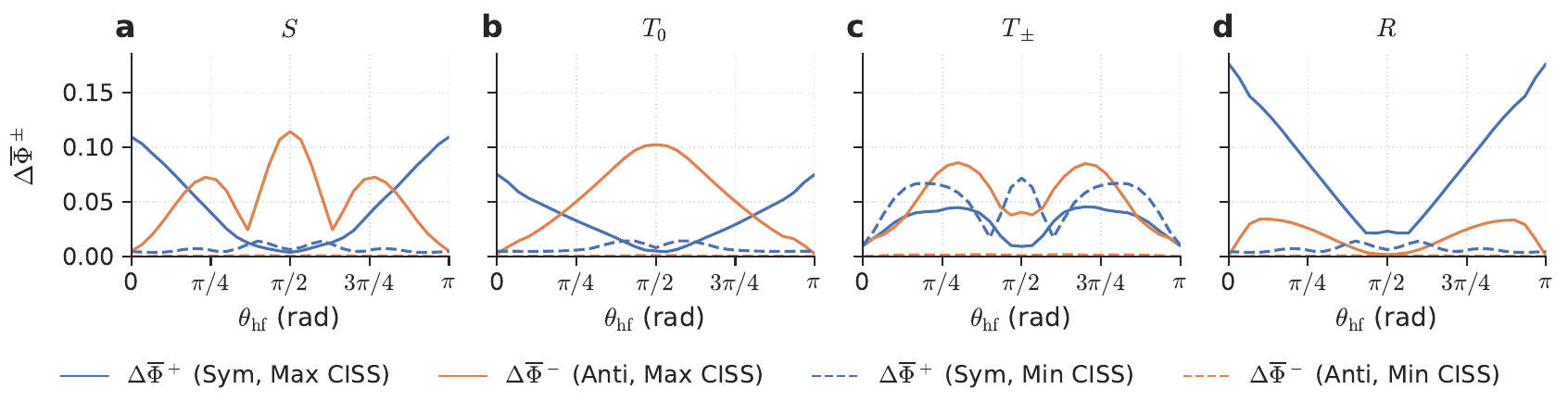}
    \caption{Symmetric ($\Delta\overline{\Phi}^+$) and antisymmetric ($\Delta\overline{\Phi}^-$) components as a function of $\theta_\text{hf}$, for $S$, $T_0$, $T_\pm$, and $R$. Parallel alignment concentrates the response in the symmetric component while orthogonal alignment transfers it to the antisymmetric component. Antisymmetric peaks in $S$ and $T_0$ near $\theta_\text{hf} = \pi/2$ confirm that hyperfine misalignment is a specific driver of field-reversal symmetry breaking, while $R$ remains predominantly symmetric across all 
orientations.}
    \label{fig:symmetry_theta_hf}
\end{figure}

The symmetry decomposition (Fig.~\ref{fig:symmetry_theta_hf}) shows that CISS progressively redistributes the orientational response from the symmetric to the antisymmetric component as the hyperfine axis moves away from the CISS axis. The $R$ channel displays a strong V-shaped symmetric component across the angular sweep, with maximum symmetry at parallel alignment. The antisymmetric components of the $S$ and $T_0$ channels peak near $\theta_\text{hf} = \pi/2$, where the hyperfine axis is orthogonal to the CISS axis. The overall pattern is that parallel alignment concentrates the response in the symmetric component, while orthogonal alignment transfers it to the antisymmetric component.

\subsubsection*{\texorpdfstring{$T_{zz}$}{Tzz} sweep at discrete angles}

\begin{figure}[h]
    \centering
    \includegraphics[width=\linewidth]{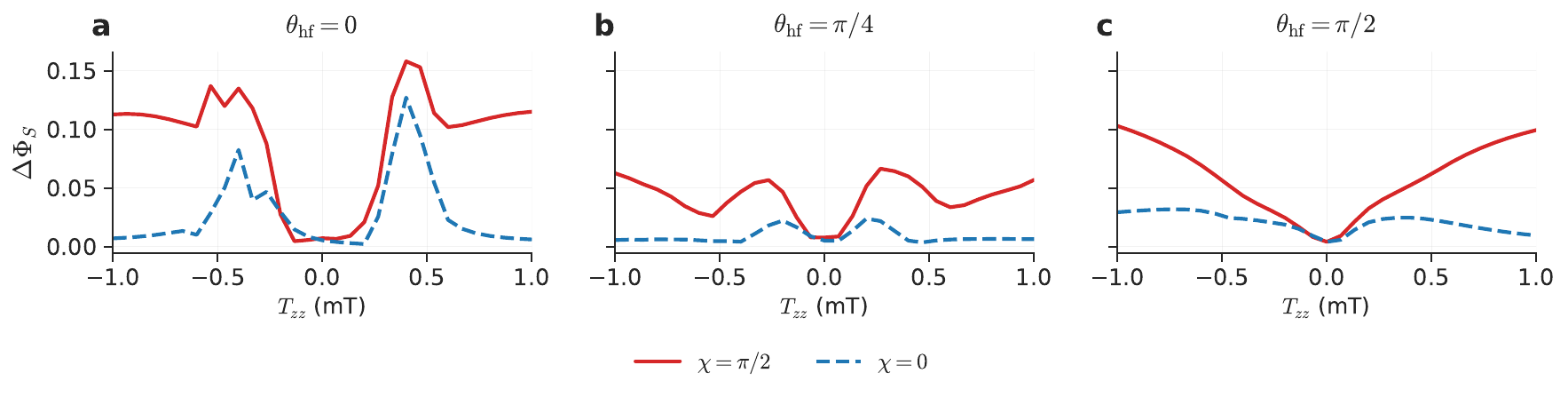}
    \caption{Singlet channel anisotropy $\Delta\Phi_S$ across a $T_{zz}$ sweep at three hyperfine angles, parallel ($\theta_\text{hf} = 0$), intermediate ($\theta_\text{hf} = \pi/4$), and orthogonal ($\theta_\text{hf} = \pi/2$). Under parallel alignment the double-peak structure is asymmetric about $T_{zz} = \SI{0}{\milli\tesla}$ and shifts to $\pm\SI{0.4}{\milli\tesla}$, consistent with the dipolar energy scale, while CISS sustains substantial anisotropy at larger $|T_{zz}|$ where the baseline collapses. The double-peak structure is recovered at the intermediate angle and replaced by a V-shaped profile under orthogonal alignment. CISS enhancement relative to the baseline is present at all three angles, with the largest CISS-to-baseline separation occurring under parallel alignment.}
    \label{fig:az_anisotropy_by_angle}
\end{figure}

To isolate how alignment conditions modify the response identified in Section~\ref{sec:aniso_hyperfine}, the $T_{zz}$ sweep is repeated at three fixed angles: $\theta_\text{hf} = 0$ (parallel), $\pi/4$ (intermediate), and $\pi/2$ (orthogonal) (Fig.~\ref{fig:az_anisotropy_by_angle}). The singlet channel is shown for clarity, as the multi-channel response becomes difficult to distinguish.

Under parallel alignment ($\theta_\text{hf} = 0$), the double-peak structure sharpens and the peak locations shift from $\pm\SI{0.27}{\milli\tesla}$ to approximately $\pm\SI{0.4}{\milli\tesla}$, consistent with the baseline dipolar coupling strength $D_0 = \SI{-0.4}{\milli\tesla}$. In contrast to the baseline RPM, where the anisotropy falls toward zero away from these peaks, CISS sustains an elevated singlet response at higher $|T_{zz}|$. The enhancement is smallest between the peaks and largest for $|T_{zz}| \gtrsim \SI{0.5}{\milli\tesla}$, where the CISS response remains roughly an order of magnitude above the baseline, reproducing the behaviour identified in Section~\ref{sec:aniso_hyperfine} where CISS maintains a growing anisotropy in the regime that the baseline RPM saturates. The response is not mirror-symmetric about $T_{zz} = \SI{0}{\milli\tesla}$ under this alignment. The positive-$T_{zz}$ peak exceeds the negative in both the baseline and CISS curves, indicating that sweeping $T_{zz}$ against the fixed transverse hyperfine and negative dipolar terms breaks the sign symmetry when the hyperfine axis is collinear with the dipolar axis. At the intermediate angle ($\theta_\text{hf} = \pi/4$), the double-peak structure near $\pm\SI{0.27}{\milli\tesla}$ is recovered, consistent with Section~\ref{sec:aniso_hyperfine}, CISS again lifts the whole curve above the near-zero baseline, and the sign asymmetry is weaker. Under orthogonal alignment ($\theta_\text{hf} = \pi/2$), the double-peak structure vanishes and is replaced by an approximately symmetric V-shaped profile centred at $T_{zz} = \SI{0}{\milli\tesla}$, with the CISS response rising monotonically. The maximum CISS-to-baseline separation is largest under parallel alignment, followed by orthogonal and then intermediate alignment, so substantial enhancement is present at all three angles rather than confined to the orthogonal case.

\begin{figure}[h]
    \centering
    \includegraphics[width=\linewidth]{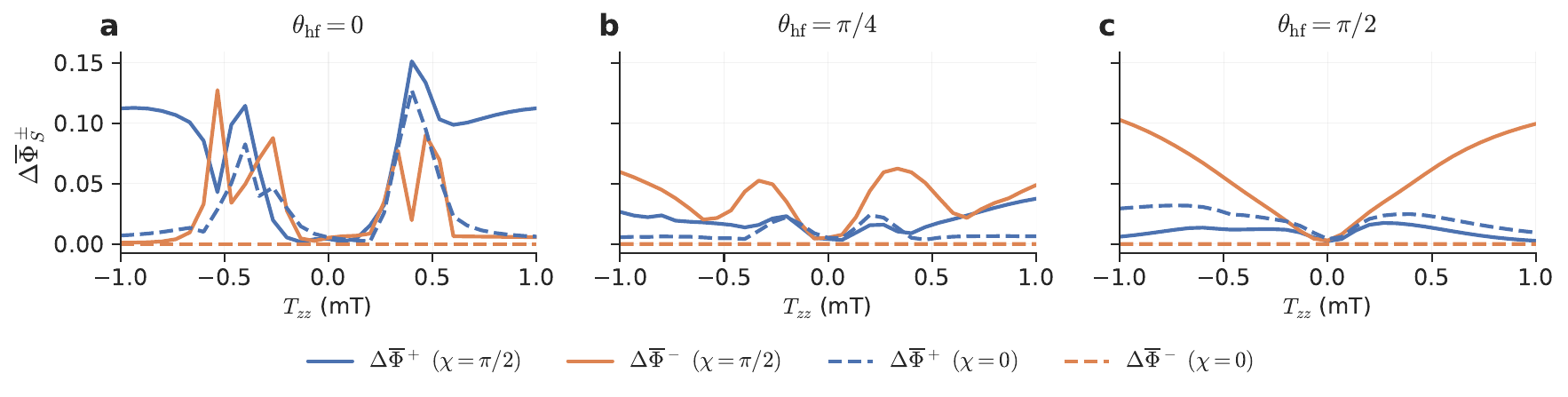}
    \caption{Symmetric ($\Delta\overline{\Phi}^{+}_S$) and antisymmetric ($\Delta\overline{\Phi}^{-}_S$) components of the singlet channel across a $T_{zz}$ sweep at three hyperfine angles, parallel ($\theta_\text{hf} = 0$), intermediate ($\theta_\text{hf} = \pi/4$), and orthogonal ($\theta_\text{hf} = \pi/2$). Under parallel alignment the response is predominantly symmetric, with the antisymmetric component negligible except for sharp localized features near $|T_{zz}| \approx |D_0|$, indicating a dipolar-mediated field-reversal asymmetry. The antisymmetric component exceeds the symmetric at the intermediate angle, and the response becomes predominantly antisymmetric under orthogonal alignment where CISS suppresses the symmetric component. The antisymmetric baseline component ($\chi = 0$) vanishes at all three angles.}
    \label{fig:az_symmetry_by_angle}
\end{figure}

The symmetric and antisymmetric decomposition at discrete angles (Fig.~\ref{fig:az_symmetry_by_angle}) follows the redistribution identified in the $\theta_\text{hf}$ sweep (Fig.~\ref{fig:symmetry_theta_hf}). Under parallel alignment ($\theta_\text{hf} = 0$) the response is predominantly symmetric, with $\Delta\overline{\Phi}^{+}_S$ reproducing the total singlet anisotropy of Fig.~\ref{fig:az_anisotropy_by_angle}. The antisymmetric component is negligible across most of the sweep and develops only sharp localized features near $|T_{zz}| \approx |D_0|$, indicating that the field-reversal asymmetry at this orientation is dipolar-mediated. At the intermediate angle ($\theta_\text{hf} = \pi/4$) the antisymmetric component exceeds the symmetric, and under orthogonal alignment ($\theta_\text{hf} = \pi/2$) the response is predominantly antisymmetric, with CISS suppressing rather than enhancing the symmetric component. The antisymmetric baseline component ($\chi = 0$) vanishes at all three angles, consistent with the absence of field-reversal asymmetry in the standard RPM. This progression from symmetric to antisymmetric dominance as the hyperfine axis rotates away from the CISS axis mirrors the $\theta_\text{hf}$ sweep.

\subsection{Dipolar Interaction}
\label{sec:dipolar}

\begin{figure}[h]
    \centering
    \includegraphics[width=\linewidth]{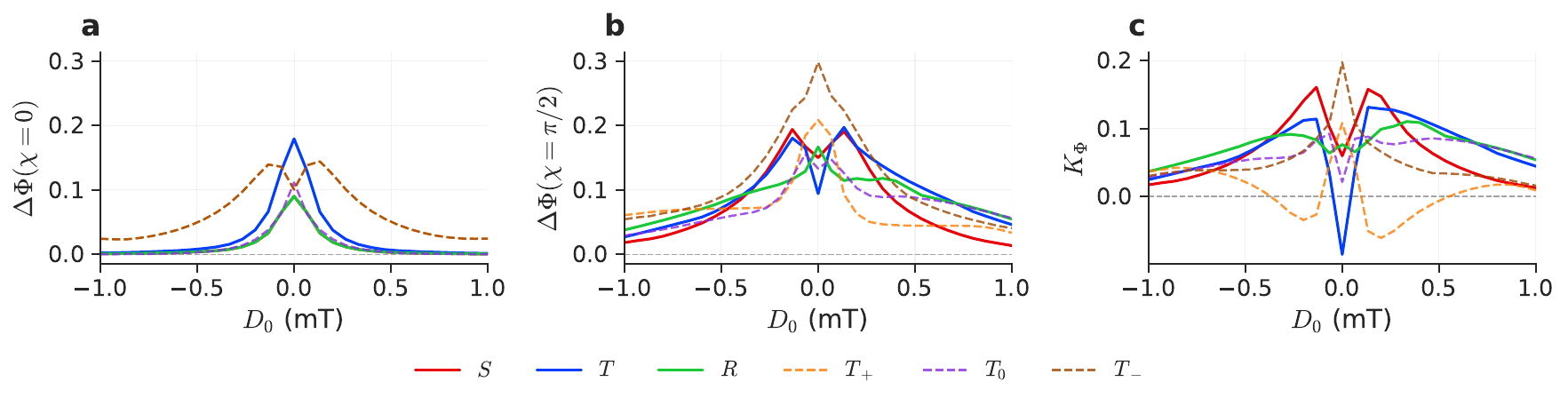}
    \caption{Orientational anisotropy and CISS-induced enhancement as a function of dipolar coupling ($D_0$). Panels (a) and (b) show $\Delta\Phi$ for the baseline RPM ($\chi = 0$) and maximal CISS ($\chi = \pi/2$) respectively. Panel (c) shows $K_\Phi$. In the baseline RPM, $S$, $T_0$, and $R$ are maximised near $D_0 = \SI{0}{\milli\tesla}$ while $T_\pm$ favour low non-zero values. Under CISS, this is reshaped with $K_\Phi$ showing negative values for the total triplet near $D_0 = \SI{0}{\milli\tesla}$, indicating CISS-induced suppression at this point.}
    \label{fig:anisotropy_D}
\end{figure}

The electron-electron dipolar interaction ($D_0$) is typically associated with quenching of orientational sensitivity in the standard RPM, as increasing dipolar splitting energetically separates the triplet sublevels from the singlet manifold and suppresses spin mixing~\cite{efimova2008role}. This is clearly reflected in the baseline RPM anisotropy (Fig.~\ref{fig:anisotropy_D}a), where the $S$, $T_0$, and $R$ channels each display a maximum near $D_0 = \SI{0}{\milli\tesla}$ before declining with increasing dipolar strength. The outer triplet channels $T_\pm$ show a qualitatively different response: their anisotropy dips at $D_0 = \SI{0}{\milli\tesla}$ and instead reaches maxima at low but non-zero values of $|D_0|$, suggesting that the $m_s = \pm 1$ states require a finite dipolar field to generate orientational contrast. Beyond $|D_0| \gtrsim \SI{0.5}{\milli\tesla}$, the dipolar interaction significantly suppresses orientational sensitivity across all channels.

Under strong CISS (Fig.~\ref{fig:anisotropy_D}b), the overall pattern is preserved but reshaped, with channel-dependent behaviour. The $S$ and $T_0$ channels shift from a maximum at $D_0 = \SI{0}{\milli\tesla}$ in the baseline RPM to a local minimum under CISS, favouring low but non-zero values of $|D_0|$. The outer triplet channels $T_\pm$ show the opposite transition, inverting from a double-peaked structure with a minimum at $D_0 = \SI{0}{\milli\tesla}$ in the baseline RPM to a single peak at $D_0 = \SI{0}{\milli\tesla}$ under CISS. The $R$ channel is the sole exception, retaining its peak at $D_0 = \SI{0}{\milli\tesla}$ in both cases. The response remains broadly symmetric across positive and negative $D_0$ values. CISS partially counteracts the quenching effect of the dipolar interaction, sustaining orientational sensitivity over a broader range of $|D_0|$ than the baseline RPM.

The chiral enhancement metric $K_\Phi$ (Fig.~\ref{fig:anisotropy_D}c) reveals that this partial counteraction is channel-dependent. For the $S$ channel, $K_\Phi$ is positive but sharply reduced near $D_0 = \SI{0}{\milli\tesla}$. The total triplet $T$ shows negative $K_\Phi$ near zero dipolar coupling, indicating that CISS actively reduces the total triplet anisotropy at this point.

\begin{figure}[h]
    \centering
    \includegraphics[width=\linewidth]{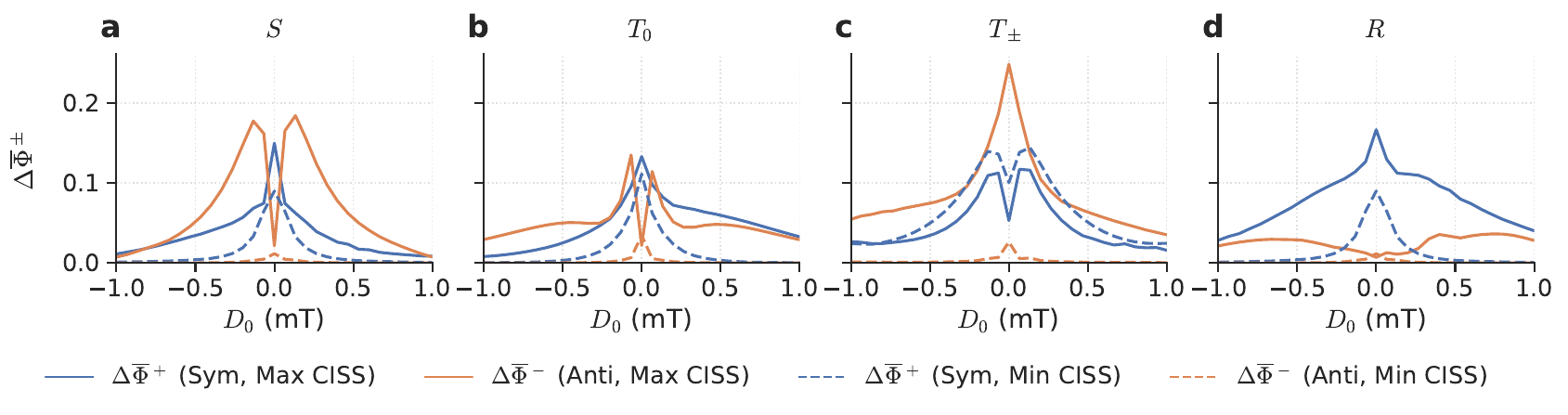}
    \caption{Symmetric ($\Delta\overline{\Phi}^+$) and antisymmetric ($\Delta\overline{\Phi}^-$) components as a function of dipolar coupling ($D_0$), for $S$, $T_0$, $T_\pm$, and $R$. The antisymmetric component under strong CISS shows a pronounced dip near $D_0 = \SI{0}{\milli\tesla}$ in $S$ and $T_0$, with corresponding symmetric maxima. Small antisymmetric components are present in the baseline RPM near $D_0 = \SI{0}{\milli\tesla}$, but CISS does not amplify these — instead the CISS antisymmetric component dips at the same point, suggesting a competing rather than cooperative mechanism.}
    \label{fig:symmetry_D}
\end{figure}

The symmetric and antisymmetric decomposition (Fig.~\ref{fig:symmetry_D}) reveals a consistent pattern centred on $D_0 = \SI{0}{\milli\tesla}$. For $S$ and $T_0$, the antisymmetric component dips, under CISS, at $D_0 = \SI{0}{\milli\tesla}$ while the symmetric component peaks at the same point. The outer triplet channels show the inverse, with the symmetric component dipping and the antisymmetric peaking at low $|D_0|$. The $R$ channel displays a sharp symmetric peak and corresponding antisymmetric minimum at $D_0 = \SI{0}{\milli\tesla}$. Notably, small non-zero antisymmetric components are present in the baseline RPM near $D_0 = \SI{0}{\milli\tesla}$ across all channels, indicating a dipolar contribution to field-reversal asymmetry independent of CISS. The CISS antisymmetric component dips at precisely this point in $S$ and $T_0$, suggesting a competing rather than cooperative relationship between dipolar-induced and CISS-induced symmetry breaking.

\subsection{Exchange Interaction}
\label{sec:exchange}

\begin{figure}[h]
    \centering
    \includegraphics[width=\linewidth]{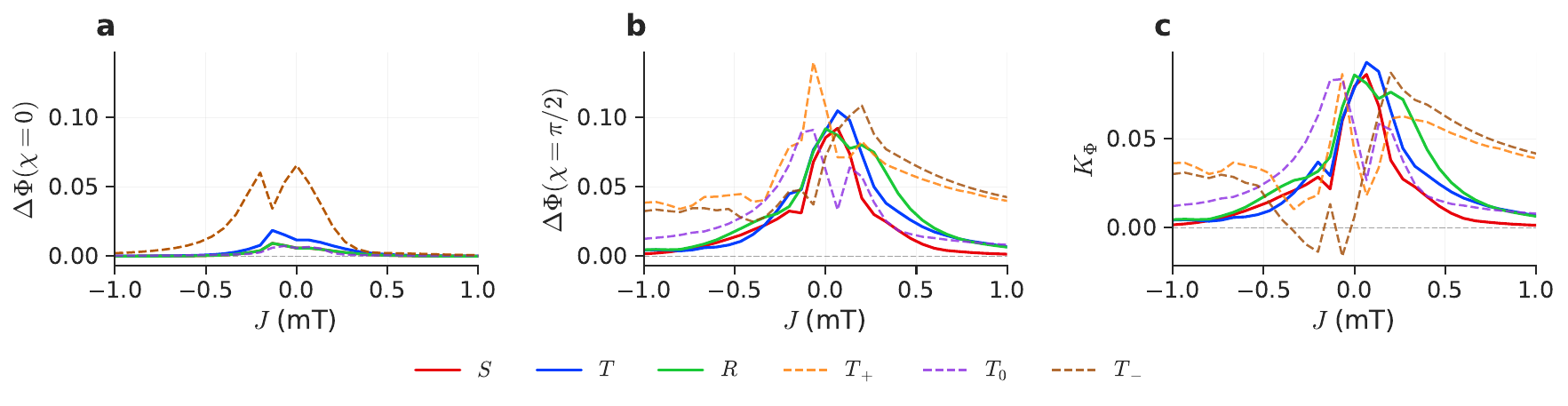}
    \caption{Orientational anisotropy and CISS-induced enhancement as a function of exchange coupling ($J$). Panels (a) and (b) show $\Delta\Phi$ for the baseline RPM ($\chi = 0$) and maximal CISS ($\chi = \pi/2$) respectively. Panel (c) shows $K_\Phi$. In the baseline RPM the anisotropy increases near $J = \SI{0}{\milli\tesla}$ and is dominated by the outer triplet channels. Under CISS the anisotropy is amplified across all channels within the same low-$|J|$ region, and $K_\Phi$ is strongly peaked at small positive $J$ rather than flat.}
    \label{fig:anisotropy_J}
\end{figure}

The isotropic exchange interaction ($J$) introduces no anisotropy tensor of its own, acting solely as a scalar shift of the singlet-triplet energy gap~\cite{lewis2018spin, efimova2008role}. Its influence on the orientational response is therefore indirect, tuning the system through the conditions under which the anisotropic hyperfine and dipolar interactions mix the singlet and triplet manifolds. In the baseline RPM (Fig.~\ref{fig:anisotropy_J}a) we observe a peaked structure concentrated near $J = \SI{0}{\milli\tesla}$. The outer triplet channels $T_\pm$ dominate, displaying a double-peaked profile with maxima near $J \approx \SI{-0.2}{\milli\tesla}$ and $\SI{0}{\milli\tesla}$, while the total triplet $T$ shows a weaker peak and the $S$, $T_0$, and $R$ channels remain near zero. Beyond $|J| \gtrsim \SI{0.4}{\milli\tesla}$ the anisotropy is quenched across all channels as the enlarged singlet-triplet gap suppresses spin mixing.

Under strong CISS (Fig.~\ref{fig:anisotropy_J}b) the degeneracy between $T_+$ and $T_-$ is lifted and the anisotropy is amplified across all channels within the same low-$|J|$ region. The peak response is centred around $J = \SI{0}{\milli\tesla}$, with the outer triplet $T_+$ reaching the largest values near $J \approx \SI{-0.1}{\milli\tesla}$ and the $S$, $T$, $T_0$, and $R$ channels forming a cluster of comparable magnitude. At larger $|J|$ the CISS response decays but remains elevated above the baseline, most visibly for the outer triplet channels.

The chiral enhancement metric $K_\Phi$ (Fig.~\ref{fig:anisotropy_J}c) is strongly peaked and concentrated at small positive $J$ for the $S$, $T$, and $R$ channels. $T_-$ dips toward zero near $J \approx \SI{-0.15}{\milli\tesla}$ before recovering to a local peak near $J \approx \SI{-0.2}{\milli\tesla}$, while $T_+$ shows a comparable local peak near $J \approx \SI{-0.1}{\milli\tesla}$ and both channels display peaks near $J \approx \SI{0.3}{\milli\tesla}$. Across all channels, the enhancement decays with increasing $|J|$ but remains weakly positive at the sweep extremes.

\begin{figure}[h]
    \centering
    \includegraphics[width=\linewidth]{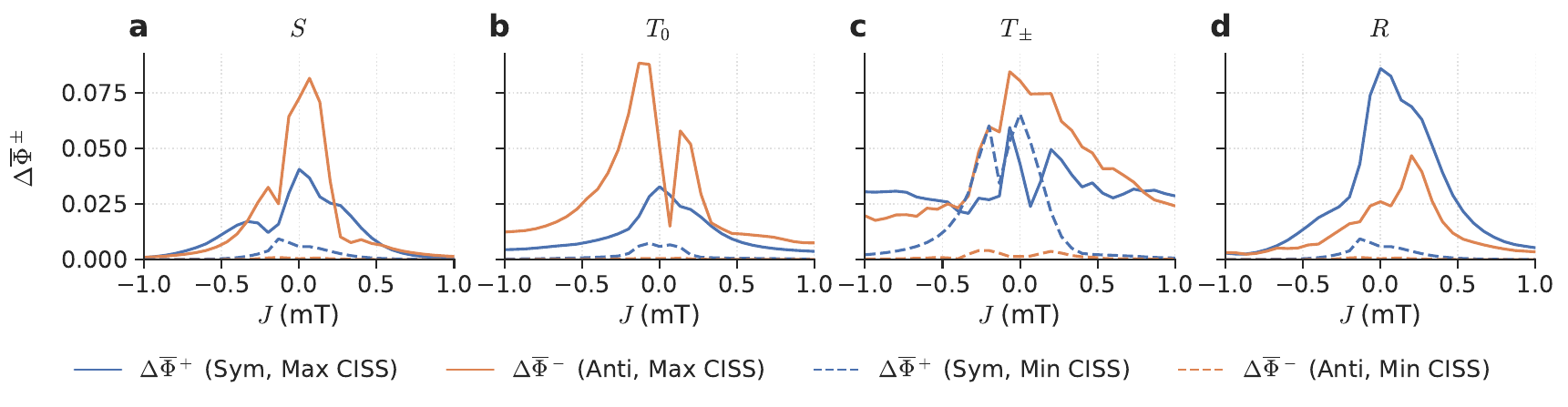}
    \caption{Symmetric ($\Delta\overline{\Phi}^+$) and antisymmetric ($\Delta\overline{\Phi}^-$) components as a function of exchange coupling ($J$), for $S$, $T_0$, $T_\pm$, and $R$. The antisymmetric component is absent in the baseline RPM and emerges under strong CISS, dominating in $S$ and $T_0$ while $R$ remains predominantly symmetric, following the same channel-dependent pattern as the dipolar and hyperfine sweeps.}
    \label{fig:symmetry_J}
\end{figure}

The symmetric and antisymmetric decomposition (Fig.~\ref{fig:symmetry_J}) confirms that the enhancement carries the same field-reversal signature identified in the other sweeps. The antisymmetric component $\Delta\overline{\Phi}^-$, absent in the baseline RPM, emerges under strong CISS with channel-dependent character. In $S$ and $T_0$ the antisymmetric component dominates, peaking near $J \approx \SI{0}{\milli\tesla}$ and $J \approx \SI{-0.1}{\milli\tesla}$ respectively. The outer triplet channels show comparable symmetric and antisymmetric contributions, while the $R$ channel remains predominantly symmetric with a sharp peak near $J \approx \SI{0}{\milli\tesla}$. This mirrors the dipolar and hyperfine sweeps, in which $R$ retains a symmetric character while $S$ and $T_0$ develop the strongest antisymmetric response.

The concentration of enhancement in the low-$|J|$ regime, where the exchange coupling is comparable to the dipolar and hyperfine energy scales, is consistent with the partial Hamiltonian cancellation between $J$ and $D_0$ shown to rescue magnetic sensitivity from dipolar quenching~\cite{efimova2008role}. As CISS enhances the same singlet-triplet mixing pathway that the exchange-tuned gap controls, its effect is concentrated where that gap is smallest and mixing is most efficient, rather than acting as an independent additive contribution.

\subsection{Two-Nucleus Model}

To assess the robustness of the single-nucleus results, the model was extended to include a second spin-$\frac{1}{2}$ nucleus coupled to the acceptor radical, with its hyperfine axis fixed at the baseline orientation. Across all parameter sweeps the antisymmetric component $\Delta\overline{\Phi}^-$ is essentially absent under both baseline and strong CISS conditions (Appendix~\ref{app:two_nuc}, Figs.~\ref{fig:sym_2nuc_B}--\ref{fig:sym_2nuc_J}). The overall anisotropy magnitudes are substantially reduced relative to the single-nucleus case, and $K_\Phi$ approaches zero or becomes negative across most of the parameter space, indicating that the CISS-induced enhancements identified in the single-nucleus model do not survive the addition of a second collinearly oriented nucleus.

\begin{figure}[h]
    \centering
    \includegraphics[width=\linewidth]{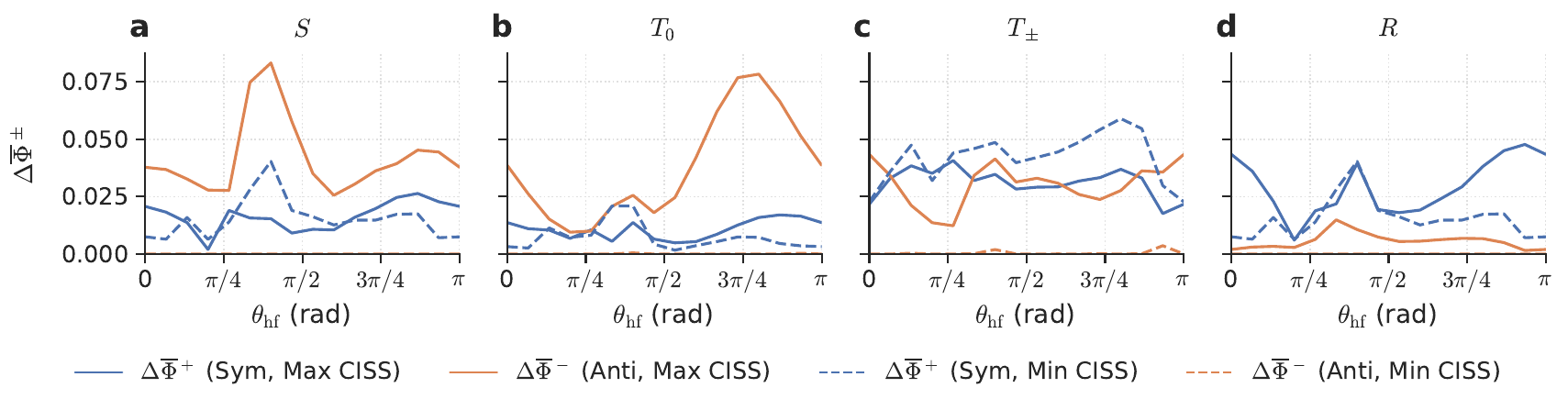}
    \caption{Symmetric ($\Delta\overline{\Phi}^+$) and antisymmetric ($\Delta\overline{\Phi}^-$) components as a function of $\theta_\text{hf}$ for the two-nucleus model, shown for $S$, $T_0$, $T_\pm$, and $R$. Unlike all other two-nucleus parameter sweeps, the antisymmetric component survives under strong CISS, confirming that mutual hyperfine misalignment is sufficient to drive field-reversal symmetry breaking even with an additional nucleus. The $S$ channel shows a broad antisymmetric peak near $\theta_\text{hf} \approx \pi/3$ and the $T_0$ channel near $3\pi/4$, though the limited angular resolution of the two-nucleus computation makes finer structure difficult to resolve reliably.}
    \label{fig:sym_2nuc_theta}
\end{figure}

The sole exception is the hyperfine axis rotation sweep (Fig.~\ref{fig:sym_2nuc_theta}). When the donor hyperfine axis is rotated while the acceptor hyperfine remains fixed, the two tensors become mutually misaligned. Under these conditions, the antisymmetric component re-emerges, confirming that genuine CISS-induced field-reversal symmetry breaking requires non-collinear internal magnetic interactions. This finding directly corroborates the geometric dependence identified in the single-nucleus results. It is the relative orientation of the internal magnetic tensors, not chiral state preparation alone, that governs whether CISS-induced symmetry breaking can manifest.

These results indicate that the conditions for CISS enhancement are more stringent than the single-nucleus model suggests. A comprehensive characterisation of the two-nucleus parameter space, including independent variation of both hyperfine orientations and strengths, represents a natural and necessary extension of the present work.
\section{Discussion and Conclusions}
\label{sec:discussion}

The systematic parametric evaluation presented here demonstrates that CISS does not function as a generic amplifier of magnetic sensitivity within the radical pair mechanism. Rather, its effects are conditional, geometry-dependent, and in many cases channel-specific. This conclusion is strengthened by extending the analysis beyond the signalling and recombination channels to the full set of spin-selective yield channels, revealing an intricate interplay in the spin dynamics that is obscured by scalar yield averages alone.

The fundamental origin of the field-reversal symmetry breaking observed throughout the parameter sweeps can be understood through the parity of the chiral initial state. In the standard RPM, the purely singlet initial state possesses even parity under spatial inversion of the external field, guaranteeing the cross-symmetry $\Phi_{T_+}(\mathbf{B}) = \Phi_{T_-}(-\mathbf{B})$~\cite{cintolesi2003anisotropic}. The chiral initial state introduced by the Luo-Hore framework modifies the density matrix to include off-diagonal coherences $|S\rangle\langle T_0| + |T_0\rangle\langle S|$, which possess odd parity under the spin inversion operator $\hat{\Pi}$ such that $\hat{\Pi}|S\rangle\langle T_0|\hat{\Pi}^{-1} = -|S\rangle\langle T_0|$~\cite{lewis2018spin}. Because the resulting chiral density matrix is not invariant under field reversal, evolving the same initial state under an inverted field alters the effective interaction geometry. This drives the macroscopic field-reversal asymmetry quantified by $\Delta\overline{\Phi}^-$ throughout the parameter sweeps.

However, the observation that this symmetry breaking collapses under a second collinearly oriented nucleus indicates that the parity argument alone is insufficient to predict when CISS enhancement will manifest. Genuine, robust field-reversal asymmetry requires non-collinear internal magnetic interactions. The single-nucleus topology lacks a hyperfine spin-flip mechanism on the acceptor electron, producing an inherently asymmetric spin-mixing pathway. Adding a second hyperfine interaction on the acceptor restores spin-flipping on both sides of the radical pair and recovers a more symmetric response. When the two hyperfine tensors are aligned, the antisymmetric component is essentially absent across all parameter sweeps despite the odd parity of the chiral initial state. It is only when the tensors are explicitly misaligned that a significant antisymmetric response survives. This identifies the relative orientation of internal magnetic tensors as the dominant geometric constraint on CISS-induced effects, emerging as the primary constraint within the explored parameter space. The intrinsic field-reversal asymmetry noted in the $T_\pm$ channels of the single-nucleus Zeeman sweep is itself a manifestation of this nuclear-topology dependence, vanishing entirely once a second collinear nucleus is introduced.

The sign dependence of the isotropic hyperfine and dipolar interactions nevertheless plays a secondary but important role. The two-dimensional parameter map shows that at low interaction strengths, the baseline RPM favours opposite-sign combinations of $a_0$ and $D_0$ for maximum anisotropy, whereas CISS shifts this toward same-sign combinations, with enhancement concentrated along the positive diagonal and greatest where $D_0 \approx \SI{0}{\milli\tesla}$. The channel-resolved analysis reveals further structure: CISS does not uniformly enhance all channels but redistributes the orientational response between symmetric and antisymmetric sectors in a channel-dependent manner, suppressing the outer triplet anisotropy in parameter regimes where it amplifies the singlet and recombination channels.

To contextualise these geometric constraints in terms of physical radical pairs, the dipolar coupling $D_0$ is strictly negative for any physically realisable inter-radical geometry, as it is determined entirely by the inter-radical distance~\cite{efimova2008role}. The positive $D_0$ regime explored here is therefore unphysical but included for completeness of the parameter landscape. The isotropic hyperfine coupling $a_0$ arises from the Fermi contact interaction and its sign depends on the electronic structure of the radical at the nuclear site~\cite{carrington1967introduction}. The finding that same-sign $a_0$ and $D_0$ favour CISS enhancement therefore implies, for physical radical pairs where $D_0 < 0$, that negative isotropic hyperfine values constitute the relevant enhancement regime within this model. The hyperfine orientation angle $\theta_\text{hf}$ is determined by the relative orientation of the hyperfine principal axes with respect to the shared dipolar and CISS reference frame imposed by the molecular geometry. The identification of non-collinear hyperfine tensors as the primary driver of robust CISS-induced symmetry breaking therefore implies a specific structural requirement on the host protein geometry, whose realisation in candidate magnetoreceptor systems remains an open question. The local protein environment has been shown to exert significant influence over hyperfine anisotropy, suggesting a structural pathway by which such geometric configurations could be realised in biological systems~\cite{adams2026quantum}.

It is not sufficient to state that CISS improves radical pair magnetic sensitivity generically. Under specific alignment conditions CISS can amplify antisymmetric field-reversal effects and increase channel-specific yields, but outside these conditions the enhancement is absent or negative. Any functional physiological role for CISS in biological magnetoreception would require the host protein to maintain a highly ordered and geometrically specific arrangement of internal magnetic interactions, with structural disorder or molecular flexibility expected to attenuate or eliminate these effects. A comprehensive characterisation of the multi-nuclear parameter space, including independent variation of both hyperfine orientations and strengths, represents the natural next step toward assessing whether these geometric conditions can be sustained in biologically relevant systems~\cite{galvan2024isotope}.
\section*{Acknowledgements}

The authors would like to thank Abbas Hassasfar for valuable discussions regarding radical pair simulations and Tristen Gwynn acknowledges support from the Stellenbosch University Postgraduate Scholarship Programme (PSP) during the course of this research.

\bibliography{ref}

@article{aiello2022chirality,
  title={A Chirality-Based Quantum Leap},
  author={Aiello, Clarice D and Abendroth, John M and Abbas, Muneer and Afanasev, Andrei and Agarwal, Shivang and Banerjee, Amartya S and Beratan, David N and Belling, Jason N and Berche, Bertrand and Botana, Antia and others},
  journal={ACS Nano},
  volume={16},
  number={4},
  pages={4989--5035},
  year={2022},
  publisher={ACS Publications},
  doi={10.1021/acsnano.1c01347}
}

@article{adams2026quantum,
  author={Adams, B. and Hassasfar, A. and Sinayskiy, I. and Nunn, A. and Guy, G. and Petruccione, F.},
  title={Quantum effects in evolution: terrestrial fine-tuning of magnetic parameters},
  journal={Computational and Structural Biotechnology Journal},
  year= {2026},
  note={arXiv:2411.03316}
}

@article{bezchastnov2026arrangement,
  title={The arrangement of anisotropic spin couplings can optimize sensitivity of the cryptochrome radical pair to the direction of geomagnetic field},
  author={Bezchastnov, Victor and Domratcheva, Tatiana},
  journal={Sci. Rep.},
  volume={16},
  number={1},
  pages={1961},
  year={2026},
  publisher={Nature Publishing Group},
  doi={10.1038/s41598-025-32180-x}
}

@article{bloom2024chiral,
  title={Chiral induced spin selectivity},
  author={Bloom, Brian P and Paltiel, Yossi and Naaman, Ron and Waldeck, David H},
  journal={Chem. Rev.},
  volume={124},
  number={4},
  pages={1950--1991},
  year={2024},
  publisher={ACS Publications},
  doi={10.1021/acs.chemrev.3c00661}
}

@book{carrington1967introduction,
  title={Introduction to Magnetic Resonance: With Applications to Chemistry and Chemical Physics},
  author={Carrington, Alan and McLachlan, Andrew D.},
  year={1967},
  publisher={Harper \& Row},
  address={New York},
  series={Harper's Chemistry Series}
}

@article{cintolesi2003anisotropic,
  title={Anisotropic recombination of an immobilized photoinduced radical pair in a 50-$\mu${T} magnetic field: a model avian photomagnetoreceptor},
  author={Cintolesi, F and Ritz, T and Kay, CWM and Timmel, CR and Hore, PJ},
  journal={Chem. Phys.},
  volume={294},
  number={3},
  pages={385--399},
  year={2003},
  publisher={Elsevier},
  doi={10.1016/S0301-0104(03)00320-3}
}

@article{dalum2019theory,
  title={Theory of chiral induced spin selectivity},
  author={Dalum, Sakse and Hedeg{\aa}rd, Per},
  journal={Nano Lett.},
  volume={19},
  number={8},
  pages={5253--5259},
  year={2019},
  publisher={ACS Publications},
  doi={10.1021/acs.nanolett.9b01707}
}

@article{efimova2008role,
  title={Role of exchange and dipolar interactions in the radical pair model of the avian magnetic compass},
  author={Efimova, Olga and Hore, PJ},
  journal={Biophys. J.},
  volume={94},
  number={5},
  pages={1565--1574},
  year={2008},
  publisher={Elsevier},
  doi={10.1529/biophysj.107.119362}
}

@article{evers2022theory,
  title={Theory of chirality induced spin selectivity: Progress and challenges},
  author={Evers, Ferdinand and Aharony, Amnon and Bar-Gill, Nir and Entin-Wohlman, Ora and Hedeg{\aa}rd, Per and Hod, Oded and Jelinek, Pavel and Kamieniarz, Grzegorz and Lemeshko, Mikhail and Michaeli, Karen and others},
  journal={Adv. Mater.},
  volume={34},
  number={13},
  pages={2106629},
  year={2022},
  publisher={Wiley Online Library},
  doi={10.1002/adma.202106629}
}

@article{fay2021chirality,
  title={Chirality-induced spin coherence in electron transfer reactions},
  author={Fay, Thomas P},
  journal={J. Phys. Chem. Lett.},
  volume={12},
  number={5},
  pages={1407--1412},
  year={2021},
  publisher={ACS Publications},
  doi={10.1021/acs.jpclett.1c00009}
}

@article{fransson2022chiral,
  title={The chiral induced spin selectivity effect what it is, what it is not, and why it matters},
  author={Fransson, Jonas},
  journal={Isr. J. Chem.},
  volume={62},
  number={11-12},
  pages={e202200046},
  year={2022},
  publisher={Wiley Online Library},
  doi={10.1002/ijch.202200046}
}

@article{gauger2011sustained,
  title={Sustained quantum coherence and entanglement in the avian compass},
  author={Gauger, Erik M and Rieper, Elisabeth and Morton, John JL and Benjamin, Simon C and Vedral, Vlatko},
  journal={Phys. Rev. Lett.},
  volume={106},
  number={4},
  pages={040503},
  year={2011},
  publisher={APS},
  doi={10.1103/PhysRevLett.106.040503}
}

@article{galvan2024isotope,
author = {Galván, Ismael and Hassasfar, Abbas and Adams, Betony and Petruccione, Francesco},
title = {Isotope effects on radical pair performance in cryptochrome: A new hypothesis for the evolution of animal migration},
journal = {BioEssays},
volume = {46},
number = {1},
pages = {2300152},
doi = {https://doi.org/10.1002/bies.202300152},
year = {2024}
}

@misc{gwynn2026ciss,
author = {Gwynn, Tristen},
title = {ciss-rpm-sweep: Simulation suite for investigating the influence of chiral state preparation on radical pair spin dynamics},
 year = {2026},
 publisher = {Zenodo},
 version = {v1.0.0},
doi = {10.5281/zenodo.20313561}
}

@article{haberkorn1976density,
  title={Density matrix description of spin-selective radical pair reactions},
  author={Haberkorn, R},
  journal={Mol. Phys.},
  volume={32},
  number={5},
  pages={1491--1493},
  year={1976},
  publisher={Taylor \& Francis},
  doi={10.1080/00268977600102851}
}

@article{hiscock2016quantum,
  title={The quantum needle of the avian magnetic compass},
  author={Hiscock, Hamish G and Worster, Susannah and Kattnig, Daniel R and Steers, Charlotte and Jin, Ye and Manolopoulos, David E and Mouritsen, Henrik and Hore, Peter J},
  journal={Proc. Natl. Acad. Sci. U.S.A.},
  volume={113},
  number={17},
  pages={4634--4639},
  year={2016},
  publisher={National Acad Sciences},
  doi={10.1073/pnas.1600341113}
}

@article{hore2016radical,
  title={The radical-pair mechanism of magnetoreception},
  author={Hore, Peter J and Mouritsen, Henrik},
  journal={Annu. Rev. Biophys.},
  volume={45},
  pages={299--344},
  year={2016},
  publisher={Annual Reviews},
  doi={10.1146/annurev-biophys-032116-094545}
}

@article{ivanov2010consistent,
  title={Consistent treatment of spin-selective recombination of a radical pair confirms the {Haberkorn} approach},
  author={Ivanov, Konstantin L and Petrova, Marina V and Lukzen, Nikita N and Maeda, Kiminori},
  journal={J. Phys. Chem. A},
  volume={114},
  number={35},
  pages={9447--9455},
  year={2010},
  publisher={ACS Publications},
  doi={10.1021/jp1048265}
}

@article{johansson2012qutip,
  title={{QuTiP}: An open-source Python framework for the dynamics of open quantum systems},
  author={Johansson, J Robert and Nation, Paul D and Nori, Franco},
  journal={Comput. Phys. Commun.},
  volume={183},
  number={8},
  pages={1760--1772},
  year={2012},
  publisher={Elsevier},
  doi={10.1016/j.cpc.2012.02.021}
}

@article{latawiec2025detecting,
  title={Detecting chirality-induced spin selectivity in chromophore-linked {DNA} hairpins using photogenerated radical pairs},
  author={Latawiec, Elisabeth I and Chiesa, Alessandro and Qiu, Yunfan and Tcyrulnikov, Nikolai A and Young, Ryan M and Carretta, Stefano and Krzyaniak, Matthew D and Wasielewski, Michael R},
  journal={Proc. Natl. Acad. Sci. U.S.A.},
  volume={122},
  number={32},
  pages={e2515120122},
  year={2025},
  publisher={National Academy of Sciences},
  doi={10.1073/pnas.2515120122}
}

@book{lewis2018spin,
  title={Spin Dynamics in Radical Pairs},
  author={Lewis, Alan},
  year={2018},
  publisher={Springer},
  doi={10.1007/978-3-030-00686-0}
}

@article{luo2021chiral,
  title={Chiral-induced spin selectivity in the formation and recombination of radical pairs: cryptochrome magnetoreception and {EPR} detection},
  author={Luo, Jiate and Hore, PJ},
  journal={New J. Phys.},
  volume={23},
  number={4},
  pages={043032},
  year={2021},
  publisher={IOP Publishing},
  doi={10.1088/1367-2630/abed0b}
}

@article{naaman2019chiral,
  title={Chiral molecules and the electron spin},
  author={Naaman, Ron and Paltiel, Yossi and Waldeck, David H},
  journal={Nat. Rev. Chem.},
  volume={3},
  number={4},
  pages={250--260},
  year={2019},
  publisher={Nature Publishing Group},
  doi={10.1038/s41570-019-0087-1}
}

@article{naaman2020chiral,
  title={Chiral molecules and the spin selectivity effect},
  author={Naaman, Ron and Paltiel, Yossi and Waldeck, David H},
  journal={J. Phys. Chem. Lett.},
  volume={11},
  number={9},
  pages={3660--3666},
  year={2020},
  publisher={ACS Publications},
  doi={10.1021/acs.jpclett.0c00474}
}

@article{naaman2012chiral,
  title={Chiral-induced spin selectivity effect},
  author={Naaman, Ron and Waldeck, David H},
  journal={J. Phys. Chem. Lett.},
  volume={3},
  number={16},
  pages={2178--2187},
  year={2012},
  publisher={ACS Publications},
  doi={10.1021/jz300793y}
}

@article{ren2021angular,
  title={Angular precision of radical pair compass magnetoreceptors},
  author={Ren, Yi and Hiscock, Hamish G and Hore, PJ},
  journal={Biophys. J.},
  volume={120},
  number={3},
  pages={547--555},
  year={2021},
  publisher={Elsevier},
  doi={10.1016/j.bpj.2020.12.023}
}

@article{ritz2000model,
  title={A model for photoreceptor-based magnetoreception in birds},
  author={Ritz, Thorsten and Adem, Salih and Schulten, Klaus},
  journal={Biophys. J.},
  volume={78},
  number={2},
  pages={707--718},
  year={2000},
  publisher={Elsevier},
  doi={10.1016/S0006-3495(00)76629-X}
}

@article{rodgers2009chemical,
author = {Christopher T. Rodgers  and P. J. Hore },
title = {Chemical magnetoreception in birds: The radical pair mechanism},
journal = {Proceedings of the National Academy of Sciences},
volume = {106},
number = {2},
pages = {353-360},
year = {2009},
doi = {10.1073/pnas.0711968106},
}

@book{sakurai2017modern,
  title={Modern Quantum Mechanics},
  author={Sakurai, J. J. and Napolitano, Jim},
  edition={2nd},
  year={2017},
  publisher={Cambridge University Press},
  doi={10.1017/9781108499996}
}

@article{smith2025chirality,
  author = {Smith, Luke D. and Tallapudi, Sukesh and Denton, Matt C. J. and Kattnig, Daniel R.},
  title = {Chirality-bolstered quantum Zeno effect enhances radical pair-based magnetoreception},
  journal = {AVS Quantum Sci.},
  volume = {7},
  number = {3},
  pages = {032601},
  year = {2025},
  publisher = {AIP Publishing},
  doi = {10.1116/5.0277712}
}

@article{tiwari2022role,
  title={Role of chiral-induced spin selectivity in the radical pair mechanism of avian magnetoreception},
  author={Tiwari, Yash and Poonia, Vishvendra Singh},
  journal={Phys. Rev. E},
  volume={106},
  number={6},
  pages={064409},
  year={2022},
  publisher={APS},
  doi={10.1103/PhysRevE.106.064409}
}

@article{tiwari2023quantum,
  title={Quantum coherence enhancement by the chirality-induced spin selectivity effect in the radical-pair mechanism},
  author={Tiwari, Yash and Poonia, Vishvendra Singh},
  journal={Phys. Rev. A},
  volume={107},
  number={5},
  pages={052406},
  year={2023},
  publisher={American Physical Society},
  doi={10.1103/PhysRevA.107.052406}
}

@article{zollner2020insight,
  title={Insight into the origin of chiral-induced spin selectivity from a symmetry analysis of electronic transmission},
  author={Z\"ollner, Martin Sebastian and Varela, Solmar and Medina, Ernesto and Mujica, Vladimiro and Herrmann, Carmen},
  journal={J. Chem. Theory Comput.},
  volume={16},
  number={5},
  pages={2914--2929},
  year={2020},
  publisher={ACS Publications},
  doi={10.1021/acs.jctc.9b01078}
}
\appendix

\section{Two-Nucleus Symmetry Decomposition}
\label{app:two_nuc}

The following figures present the symmetric ($\Delta\overline{\Phi}^+$) and antisymmetric ($\Delta\overline{\Phi}^-$) components of the orientational yield distribution for the two-nucleus model across the Zeeman, isotropic hyperfine, anisotropic hyperfine, dipolar, and exchange parameter sweeps. All two-nucleus computations are evaluated over a reduced $11 \times 11$ angular grid. Across all sweeps the antisymmetric component is essentially absent under both baseline and strong CISS conditions, in contrast to the single-nucleus results. The hyperfine axis rotation sweep, which is the sole exception, is presented in the main text (Fig.~\ref{fig:sym_2nuc_theta}).

\begin{figure}[h]
    \centering
    \includegraphics[width=\linewidth]{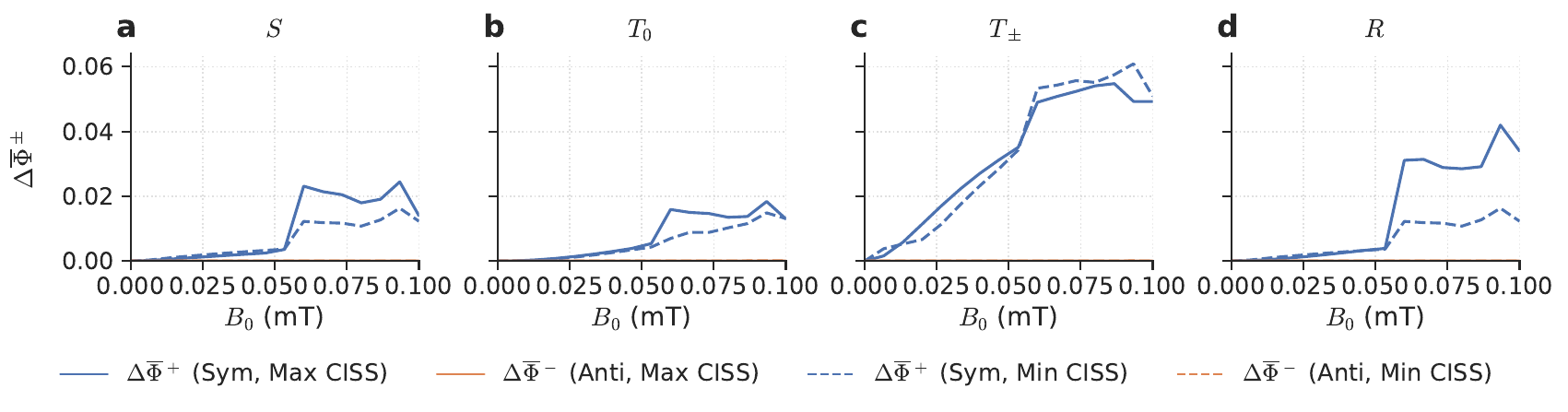}
    \caption{Symmetric ($\Delta\overline{\Phi}^+$) and antisymmetric ($\Delta\overline{\Phi}^-$) components as a function of Zeeman coupling ($B_0$) for the two-nucleus model, shown for $S$, $T_0$, $T_\pm$, and $R$. The antisymmetric component is absent under both baseline and strong CISS conditions, and the symmetric component shows only minor differences between $\chi = 0$ and $\chi = \pi/2$, indicating that CISS-induced field-reversal symmetry breaking does not survive the addition of a second collinearly oriented nucleus in this sweep.}
    \label{fig:sym_2nuc_B}
\end{figure}

\begin{figure}[h]
    \centering
    \includegraphics[width=\linewidth]{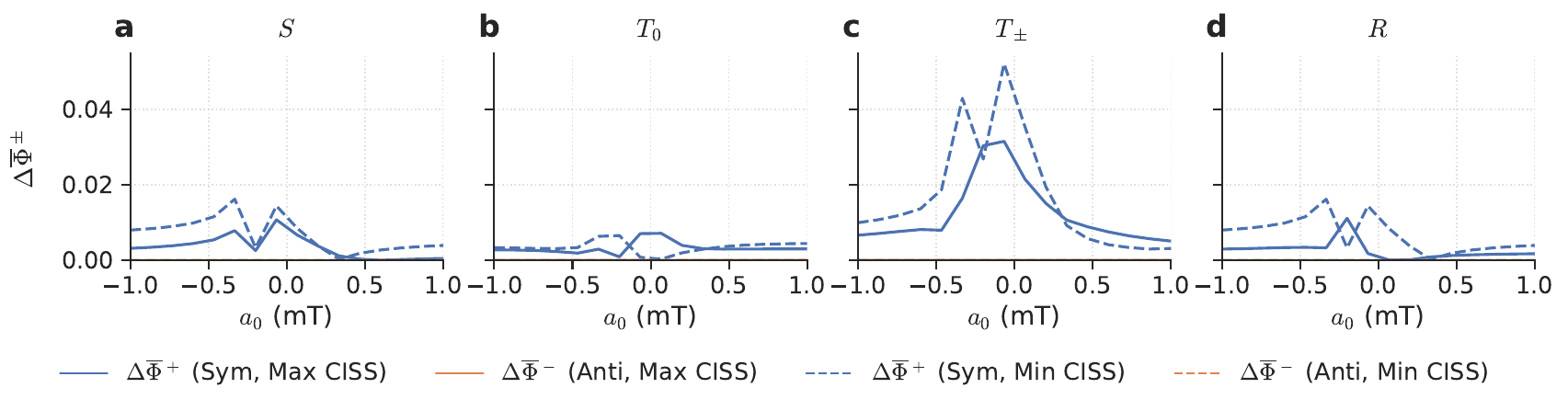}
    \caption{Symmetric ($\Delta\overline{\Phi}^+$) and antisymmetric ($\Delta\overline{\Phi}^-$) components as a function of isotropic hyperfine coupling ($a_0$) for the two-nucleus model, shown for $S$, $T_0$, $T_\pm$, and $R$. The antisymmetric component vanishes across the full $a_0$ range under both chirality conditions. The overall anisotropy magnitudes are substantially reduced relative to the single-nucleus case.}
    \label{fig:sym_2nuc_Ai}
\end{figure}

\begin{figure}[h]
    \centering
    \includegraphics[width=\linewidth]{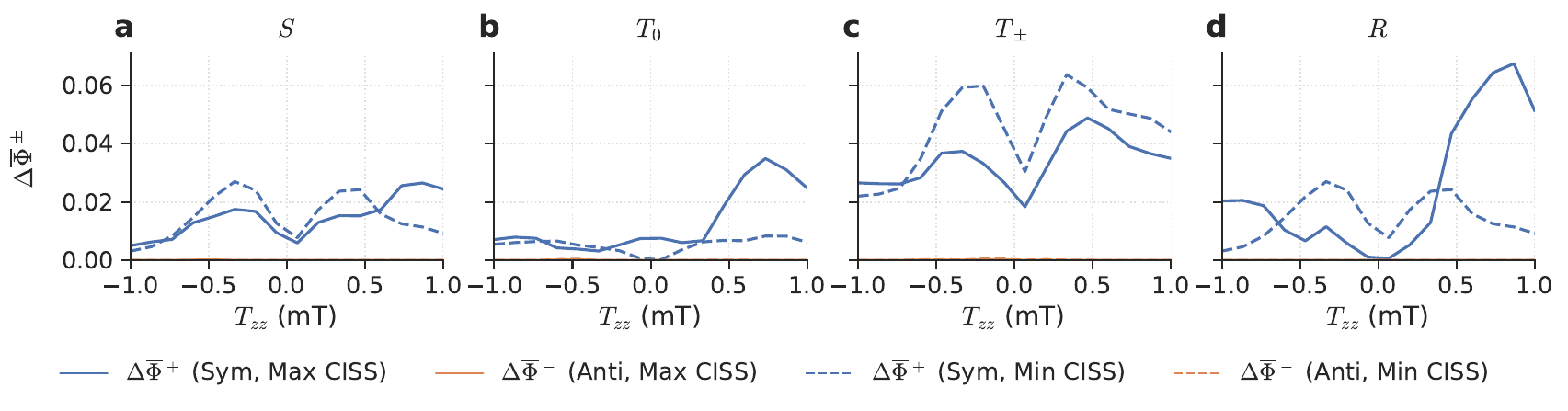}
    \caption{Symmetric ($\Delta\overline{\Phi}^+$) and antisymmetric ($\Delta\overline{\Phi}^-$) components as a function of anisotropic hyperfine coupling ($T_{zz}$) for the two-nucleus model, shown for $S$, $T_0$, $T_\pm$, and $R$. The antisymmetric component is absent throughout, and the double-lobe sensitivity structure observed in the single-nucleus case is substantially suppressed.}
    \label{fig:sym_2nuc_Az}
\end{figure}

\begin{figure}[h]
    \centering
    \includegraphics[width=\linewidth]{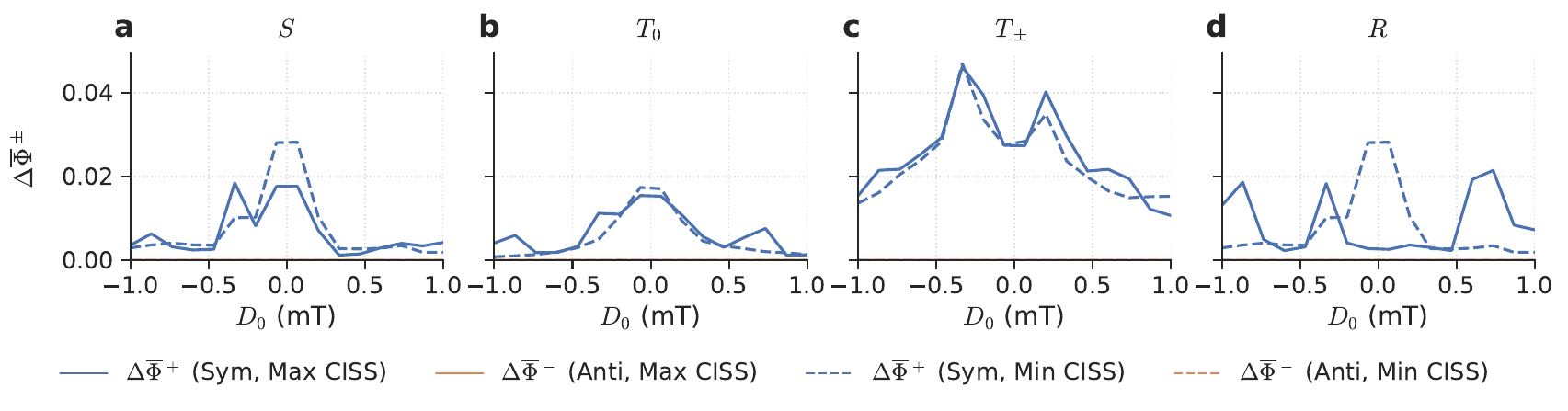}
    \caption{Symmetric ($\Delta\overline{\Phi}^+$) and antisymmetric ($\Delta\overline{\Phi}^-$) components as a function of dipolar coupling ($D_0$) for the two-nucleus model, shown for $S$, $T_0$, $T_\pm$, and $R$. The antisymmetric component is absent across the full $D_0$ range, and the channel-dependent inversion pattern observed near $D_0 = \SI{0}{\milli\tesla}$ in the single-nucleus model is not reproduced.}
    \label{fig:sym_2nuc_D}
\end{figure}

\begin{figure}[h]
    \centering
    \includegraphics[width=\linewidth]{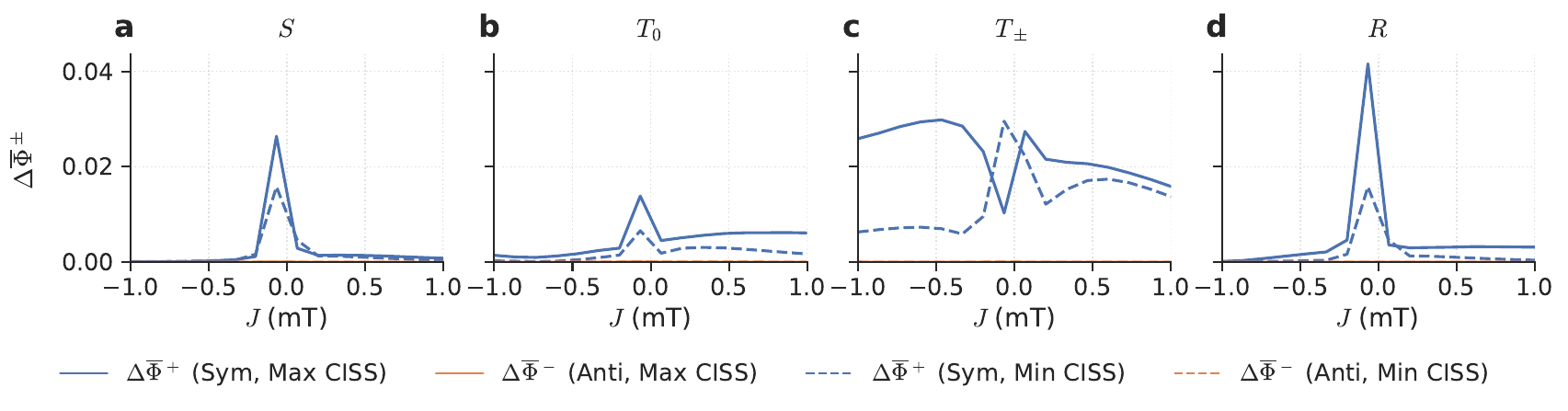}
    \caption{Symmetric ($\Delta\overline{\Phi}^+$) and antisymmetric ($\Delta\overline{\Phi}^-$) components as a function of exchange coupling ($J$) for the two-nucleus model, shown for $S$, $T_0$, $T_\pm$, and $R$. The antisymmetric component is absent across the full $J$ range under both chirality conditions, and the sharply peaked enhancement observed near $J = \SI{0}{\milli\tesla}$ in the single-nucleus model is substantially suppressed.}
    \label{fig:sym_2nuc_J}
\end{figure}

\end{document}